\documentclass[12pt]{iopart}
\usepackage{amsfonts}
\usepackage{graphicx}

\begin{document}

\title[A novel series solution]{A novel series solution to the renormalization
group equation in QCD}

\author{B A Magradze}

\address{A. Razmadze Mathematical Institute, M. Aleksidze St. 1,
Tbilisi 0193, Georgia}

\ead{magr@rmi.acnet.ge}


\begin{abstract}

Recently, the QCD renormalization group (RG) equation at higher
orders in MS-like renormalization schemes has been solved for the
running coupling as a series expansion in powers of the exact
2-loop order coupling. In this work, we prove that the power
series converges to all orders in perturbation theory. Solving the
RG equation at higher orders, we determine the running coupling as
an implicit function of the 2-loop order running coupling. Then we
analyze the singularity structure of the higher order coupling in
the complex 2-loop coupling plane. This enables us to calculate
the radii of convergence of the series solutions at the 3- and
4-loop orders as a function of the number of quark flavours
$n_{\rm f}$. In parallel, we discuss in some detail the
singularity structure of the ${\overline{\rm MS}}$ coupling at the
3- and 4-loops in the complex momentum squared plane for $ 0\leq
n_{\rm f} \leq 16 $. The correspondence  between the singularity
structure of the running coupling in the complex momentum squared
plane and the convergence radius of the series solution is
established. For sufficiently large $n_{\rm f}$ values, we find
that the series converges for all values of the momentum squared
variable $Q^2=-q^2>0$. For lower values of $n_{\rm f}$, in the
${\overline{\rm MS}}$ scheme, we determine the minimal value of
the momentum squared $Q_{\rm min}^2$ above which the series
converges. We study properties of the non-power series
corresponding to the presented  power series solution  in the QCD
Analytic Perturbation Theory approach of Shirkov and Solovtsov.
The Euclidean and Minkowskian versions of the non-power series are
found to be uniformly convergent over whole ranges of the
corresponding  momentum squared variables.
\end{abstract}

\section{Introduction}
It is known that the QCD running coupling at the 2-loop order in
MS-like (massless) renormalization schemes can be solved
explicitly as a function of the scale in terms of the Lambert W
function \cite{ggk,m1,appel}. The Lambert W function is the
multivalued solution of
\begin{equation}
\label{LW} W_{k}(z)\exp\{W_{k}(z)\}=z,
\end{equation}
the branches of $W$ are denoted $W_{k}(z)$, $k=0,\pm 1,\ldots .$
An exhaustive review of the Lambert-$W$ function may be found in
ref. \cite{lamb}. The relevant branch of $W(z)$ which is used to
determine the coupling depends on the number of light quark
flavours $n_{\rm f}$.  For a real positive momentum squared $Q^2$
\footnote[1]{Here $Q^2=-q^2=-(q^{0})^{2}+{\vec{q}}^{2}$ and
$Q^2>0$ in the Euclidean domain.} (and above the Landau
singularity if $0\leq n_{\rm f}\leq 8$) the 2-loop coupling takes
the form
\begin{equation}
\label{w2}
 \alpha_{\rm s}^{(2)}(Q^2,n_{\rm
f})=\left\{\begin{array}{ll}
-({\beta}_0/{\beta}_1)(1+W_{-1}(z_{Q}))^{-1},&\mbox{ if $0\leq n_{\rm f}\leq 8$}\\
-({\beta}_0/{\beta}_1)(1+W_{0}(z_{Q}))^{-1},&\mbox{if $9\leq
n_{\rm f}\leq 16$}\end{array}\right.
\end{equation}
where $z_{Q}=-(eb_{1})^{-1}(Q^2/\Lambda^2)^{-1/b_{1}}$,
$\beta_{0}$ and $\beta_{1}$ are the first two $\beta$-function
coefficients, $b_{1}=\beta_1/\beta_0^2$ and
$\Lambda\equiv\Lambda_{\overline {\rm MS}}$ is the conventional
${\overline {\rm MS}}$ scheme QCD parameter. Using formula
(\ref{w2}), the analytical structure of the 2-loop coupling in the
complex $Q^2$ plane was determined \cite{ggk,m1,m2}. The
motivation for these  studies was a need for the development of
dispersive methods [6-25]. The dispersive approach has been
devised to extend  properly modified perturbation theory
calculations towards the low-energy region \cite{dmw, ss,gr}. The
most simple and elaborated variant of the dispersive approach, the
Shirkov-Solovtsov Analytic Perturbation Theory (APT), was
formulated in refs.~ \cite{ss} and \cite{mss} (for a review on APT
and many original references see refs.~\cite{ss1} and \cite{prosp}). It should be
remarked that in the time-like region APT is equivalent to the
``contour improved perturbation theory" proposed previously in
ref.~ \cite{piv} (see also refs.~[27-31]). The relation between
this framework and APT was discussed in refs.~\cite{shirk2,join}.
More sophisticated nonperturbative modifications of the (minimal)
analytic QCD model of Shirkov and Solovtsov were also presented
[18-23]. A generalization of APT to non-integer (fractional)
powers of the running coupling has also been proposed and
successfully used to calculate the three-point functions in QCD
\cite{bakul}.

The 2-loop explicit solution (\ref{w2}) was soon found to have a
more important application. In ref.~ \cite{kur}, the running
coupling in the $k$-th order ($k\geq 3$) in a MS-like
renormalization scheme was expanded in powers of  the exact
two-loop order coupling (here and hereof we omit the argument
$n_{\rm f}$)
\begin{equation}
\label{ser} \alpha_{\rm
s}^{(k)}(Q^2)=\sum_{n=1}^{\infty}c_{n}^{(k)}\alpha_{s}^{(2)n}(Q^2).
\end{equation}
On this basis, author of ref.~ \cite{kur} has proposed a new
method for reducing the scheme ambiguity for QCD observables. A
similar expansion (motivated differently for an observable
depending on a single scale) was suggested in ref.~\cite{mx}. Note
that the analogical expansion but in powers of the approximate
(asymptotic) 2-loop coupling has previously been introduced in
ref.~\cite{rops}. This expansion was used to construct the running
coupling  with consistent matching conditions at the quark
thresholds in the 3-loop order. However, if the asymptotic 2-loop
coupling is used, the coefficients of the expansion depend on the
scale $Q^2$. The main advantage of (\ref{ser}) is that it allows
us to write QCD observables, in massless renormalization schemes,
as series in powers of the renormalization scheme independent
quantity, the exact explicit 2-loop coupling (\ref{w2}). One could
introduce a similar expansion in powers of the 1-loop scheme
independent coupling as well. However such a series would not be
useful, since it could not describe the double logarithmic
singularities of the higher order coupling. Recently, the series
(\ref{ser}) has been used to construct exact explicit expressions
for Euclidean and Minkowskian observables within APT \cite{join,
m4} (see also ref.~ \cite{howe}). In practice, the first few terms
in series (\ref{ser}) give the excellent approximations to the
coupling even in the infrared region \cite{join,m4}

The main purpose of this paper is to present a detailed
mathematical investigation of the series (\ref{ser}). In Sect.~2
we discuss in some details the singularity structure of the
${\overline {{\rm MS}}}$ coupling, in higher orders, in the
complex $Q^2$ plane. In particular, we determine the locations of
the Landau singularities of the coupling (at the 3- and 4-loops)
as a function of $n_{\rm f}$ for the $n_{\rm f}$ values into the
range of validity of the asymptotic freedom of QCD. A similar
investigation (but for large  $n_{\rm f}$ values when the
$\beta$-function has a positive fixed point) has previously been
undertaken by the authors of refs.~\cite{gg,gkarl} using a
different technique, whose work does not overlap the material in
Sect.~2 to a marked extent. In Sect.~3 we prove that the series
(\ref{ser}) in the MS-like renormalization schemes has a positive
radius of convergence to all orders in perturbation theory. In the
proof we use the methods of the analytical theory of differential
equations. In Sect.~4 we solve a higher-order RG equation for the
running coupling implicitly as a function of the 2-loop running
coupling. By means of the obtained transcendental equation, we
determine the analytical structure of the higher order coupling in
the complex 2-loop coupling plane. As a result, we evaluate
analytically the radii of convergence of series (\ref{ser}) at 3-
and 4-loops as a function of $n_{\rm f}$.
In Sect.~5 we determine the convergence region of the series solution with
respect to the momentum squared variable $Q^2$. For sufficiently
large $n_{\rm f}$ values ($n_{\rm f}\geq 14$ in the
${\overline{\rm MS}}$ scheme), we find that the series converges
for all $Q^2>0$.  For the lower $n_{\rm f}$ values, we determine
the minimal value $Q^2_{\rm min}$ above which the series
converges. We compare this scale, at $n_{\rm f}=3$, with the
infrared boundary of perturbative QCD estimated within two
different nonperturbative frameworks. In Sect.~6 we study
properties of the dispersive images of the series solution
(\ref{ser}), the non-power series determined in the sense of the
QCD Analytic Perturbation Theory approach of Shirkov and
Solovtsov, both in the space- and time-like regions. Our
conclusions are given in Sect.~7. In the Appendix we collect some
relevant formulas which we need in our computations.

\section{The Analytic Structure of the Coupling to Higher Orders}

In this section, we will determine the location  of the Landau
singularities of the coupling  in the complex $Q^2$ plane at the
3-loop and 4-loop orders. As we shall see, there is a close
relation between these singularities and the convergence
properties of the series (\ref{ser}). For large $n_{f}$ values
\footnote[2]{Note that QCD as a realistic theory appears only for
6 flavours and below. However, there are different theoretical
motivations to consider the multi-flavour theory for $6<n_f\leq
16$ (see, for example, \cite{banks, miran, oz}).}, we will
reproduce part of the results of \cite{gg,gkarl} using another
technique. Let us first give some familiar aspects of the RG
equation in the way we prefer to formulate it. It is more
convenient to introduce the dimensionless variable
$u=Q^2/\Lambda^2$ and a modified running coupling,
$a(u)=\beta_{0}\alpha_{\rm{s}}(Q^2)$, satisfying the RG equation
\begin{equation}
\label{coup1} u\frac{{d} a(u)}{{d} u}={\bar\beta}^{(k)}(a(u))
=-\sum_{n=0}^{k-1} b_{n} a^{n+2}(u)
\end{equation}
where ${\bar\beta}^{(k)}(a)=
{\beta}_{0}{\beta}^{(k)}(a/\beta_{0})$ and
$b_{n}=\beta_{n}/\beta_{0}^{n+1}$ (for our notations see the
Appendix). The $\overline{MS}$ scheme values of the first three
coefficients $b_{1,2,3}$  are listed in Table A.1. With this
normalization of the coupling relevant perturbative formulas in
higher order applications of renormalization group become simple
\cite{shi,AA1,piv1}. Suppose that the solution to (\ref{coup1})
$a(u)$ has a singularity at some finite point, $u=u_{\rm L}$, in
the complex $u$-plane, i.e. $ a(u)\rightarrow\infty $ as $ u
\rightarrow u_{\rm L} $. It follows then from the differential
equation (\ref{coup1}) that
\[
a^{(k)}(u)\approx(u^{(k)}_{\rm L}/(u-u_{\rm L}^{(k)}))^{1/k} \quad
\mbox{as}\quad u\rightarrow u_{\rm L}^{(k)}.
\]
However, to confirm the existence of the singularities and to
determine their positions, a detailed investigation is required in
each finite order of perturbation theory.

Let us integrate Eq.~(\ref{coup1}) for sufficiently large real
positive values of $u=\exp(t)$ in the neighborhood of a real point
$u_{0}=\exp(t_{0})$
\begin{equation}
\label{gsn} t=\ln{u}=T^{(k)}(a) \quad \mbox{where} \quad
T^{(k)}(a)=\int_{a_{0}}^{a}\{{\bar{\beta}}^{(k)}(a')\}^{-1}{d}
a'+t_{0},
\end{equation}
this can also be written
\begin{equation}
\label{gsn1} t=a^{-1}+b_{1}\ln(a)+{\tilde
T}^{(k)}(a)+\tilde{t_{0}},
\end{equation}
where ${\tilde T}^{(k)}(a)$ is a regular at zero function
\begin{equation}
\label{regf} {\tilde T}^{(k)}(a)=\int_{a_0}^{a}G^{(k)}(a')d a':
\quad G^{(k)}(a)=1/{\bar{\beta}}^{(k)}(a)+1/a^2-b_1/a,
\end{equation}
here the integration constant has been redefined:
$\tilde{t}_{0}=t_0-a_0^{-1}-b_{1}\ln(a_0)$. The conventional
definition of  the scale $\Lambda$ parameter \cite{bbdm} leads us
to the condition ${\tilde t}_0=-{\tilde T}^{(k)}(0)$
\footnote[3]{the term proportional to $1/{\ln}^2(u)$ in the
asymptotic expansion of $a(u)$ at large $u$ should be
suppressed.}. With this choice Eq.~(\ref{gsn1}) reads
\begin{equation}
\label{gsn2} t=a^{-1}+b_1\ln(a)+\int_{0}^{a}G^{(k)}(a') d a'.
\end{equation}
We could write in place of (\ref{gsn2}) another but related
formula \cite{stev}
\begin{equation}
\label{gsstev}
t=a^{-1}-b_{1}\ln(b_{1}+a^{-1})+\int_{0}^{a}g^{(k)}(a'){d}a'
\end{equation}
where
\begin{equation}
 \label{gk}
 \fl g^{(k)}(a)=
(\bar{\beta}^{(k)}(a))^{-1}-(\bar{\beta}^{(2)}(a))^{-1}
=(1+b_{1}a)^{-1}\left(\sum_{n=0}^{k-1}b_{n}a^{n}\right)^{-1}
\sum_{n=2}^{k-1}b_{n}a^{n-2},
\end{equation}
and $\bar{\beta}^{(2)}(a)$ is the 2-loop $\bar{\beta}$-function.
The function $T^{(k)}(a)$ can be expressed in terms of the
elementary functions. In the 3-loop case, we can write
\begin{equation}
\label{3lim}
T^{(3)}(a)=a^{-1}+b_{1}\ln(a)+T^{(3)}_{1}(a)-T^{(3)}_{1}(0),
\end{equation}
where
\begin{equation}
\label{3cases}  \fl T^{(3)}_{1}(a)=\left\{\begin{array}{ll}
\displaystyle{-0.5b_{1}\ln(P^{(3)}(a))+\frac{2b_{2}-b_{1}^2}{\sqrt{\Delta^{(3)}}}
\arctan\left(\frac{b_{1}+2b_{2}a}{\sqrt{\Delta^{(3)}}}\right)}&\mbox{if $0\leq n_{f}\leq 5$}\\
\displaystyle{\frac{\ln(a-a_{1})}{(a_1-a_2)(1+b_{1}a_{1})}-
\frac{\ln(a_2-a)}{(a_1-a_2)(1+b_{1}a_{2})}}&\mbox{if $6\leq
n_{f}\leq 16$,}
\end{array}
\right.
\end{equation}
$P^{(3)}(a)=b_{2}a^2+b_{1}a+1$, $\Delta^{(3)}=4b_{2}-b_{1}^2$, and
$a_{1,2}=(-b_{1}\pm\sqrt{-\Delta^{(3)}})/(2b_{2})$. In the
${\overline{\rm MS}}$ scheme $\Delta^{(3)}>0$ $(<0)$ if $0\leq
n_{f}\leq 5$ $(6\leq n_{f}\leq 16)$ (see Tables 1 and 2).

Let us specify the locations of the roots of the algebraic
equation
\begin{equation}
\label{algeq}
P^{(k)}(a)=\bar{{\beta}}^{(k)}(a)/a^{2}=-\sum_{n=0}^{k-1}b_{n}a^n=0,
\end{equation}
the non-trivial zeros of the $\bar\beta$ function, for different values of
$n_{\rm f}$. For $0\leq n_{\rm f}\leq 7$, in the
4-loop order in the $\overline{\rm MS}$ scheme equation
(\ref{algeq}) has one negative real root $a_1<0$ and a pair of
complex conjugate roots $a_2=\bar{a}_3$ (see Table 3). We will
assume that $\Im{(a_2)}<0$. Then Eq.~(\ref{gsn2}) can be written
as
\begin{equation}
\label{4loop1} t=a^{-1}+b_{1}\ln(a)+T_{1}^{(4)}(a)-T_{1}^{(4)}(0),
\end{equation}
where $T_{1}^{(4)}(a)$ is a regular at zero function
\begin{eqnarray}
\label{smalln} T_{1}^{(4)}(a)=-b_{3}^{-1}\left\{E_1\ln(a-a_1)+
\Re(E_2)\ln[(a-a_2)(a-a_3)] \right.\nonumber\\
\left.+2\Im(E_2)\arctan{\left(\frac{a-\Re(a_2)}
 {|\Im(a_2)|}\right)}\right\},
\end{eqnarray}
with
\[
E_{i}=\{a_{i}^2(a_i-a_j)(a_i-a_k)\}^{-1},\quad i\neq j\neq k
\]
and $(i,j,k)$ is a cyclic permutation of $(1,2,3)$. Equation
(\ref{smalln}) was derived for real positive values of $a$. It may
be continued analytically in the complex $a$-plane choosing the
relevant branches for each elementary function on the right hand
side.  For unphysical  values $8\leq n_f\leq 16$, in the 4-loop
case Eq.~(\ref{algeq}) has three real roots: $a_1<0$, $0<a_2<a_3$
(see Table 4). Let $a$ be positive lying in the interval
$0<a<a_2$. Formula (\ref{gsn2}) may now be rewritten
\begin{equation}
\label{bign}
t=a^{-1}+b_1\ln(a)-b_3^{-1}\sum_{i=1}^{3}E_{i}\ln(a_{i}^{-1}(a_{i}-a)).
\end{equation}
For complex values of $a$, the analytical continuation of
Eq.~(\ref{bign}) can be easily performed assuming that each
logarithm on the right of (\ref{bign}) have its principal value.

We have to determine the analytical continuation of the coupling
starting from the implicit solution (\ref{gsn}).
This implies  a preliminary  study of the analytical properties
of the inverse function $t=T^{(k)}(a)$ in the complex coupling
plane. The one-valued branch of this function may be defined in
the cut complex $a$-plane choosing the cuts carefully. The
physical branch may be determined from the requirement that the
branch yields a real positive t for real positive and sufficiently
small values of $a$. It is convenient to determine the analytical
continuation of $T^{(k)}(a)$ starting directly from the integral
representation (\ref{gsstev}). The integral there should be
regarded as a line integral in the complex $a$-plane. The line
must be deformed to avoid singularities of the integrand.
It is seen from (\ref{gsn1}) that the function $T^{(k)}(a)$  has a
simple pole as well as a logarithmic branch point at $a=0$. In
addition, there are logarithmic singularities  at the
roots of the algebraic equation (\ref{algeq}).
For $0<n_{\rm f}\leq 5$, in the 3-loop case (in the ${\overline
{{\rm MS}}}$ scheme) Eq.~(\ref{algeq}) has a pair of complex
conjugate roots, while it has two real (positive and negative)
roots for $6\leq n_{\rm f}\leq 16$ (see Tables 1 and 2).  The
4-loop case has  already been discussed above. We must make a
branch cut along the negative $a$-axis $\{a:-\infty<a<0\}$
corresponding to the logarithmic branch point at zero. In the
cases, where there is a pair of complex conjugate branch points
(say $a_{2,3}$  in the 4-loop case) we must choose additional
branch cuts. One possibility is to choose the cuts parallel to the
imaginary axis (see Fig. 1(a)), so that these branch points are
connected by the cut running through the infinity.  In the 4-loop
case, we choose the cuts
$\{a:-\infty<\Im(a)<\Im(a_2),
\Im(a_{3})<\Im(a)<\infty,\Re(a)=\Re(a_{3})\}$ (see Fig. 1(a)).  The analytical
continuation of $T^{(k)}(a)$ in the cut complex $a$-plane will be
determined uniquely if we require that $T^{(k)}(a)$ is real for a
real positive and sufficiently small values of $a$. Note that the
above considered choice for the cuts is not unique. We could, for
example, choose the cuts running along straight lines connecting
the complex conjugate branch points to the origin. Nevertheless,
former possibility (which we accept in this paper) seems to be
preferable: with this choice $t$ as a function of the phase of $a$
will be continuous in the neighborhood of $a=0$ with the exception
of the cut running along the negative $a$-axis.

Consider now the theoretical cases with only real roots. Let $a_1$
be the negative root, and $a_2$ be the positive one (in the 4-loop
case $a_2$ is the smallest positive root). Then the branch cuts
may be chosen along the real intervals $\{a:-\infty<a<0\}$ and
$\{a: a_2<a<\infty\}$ (see Fig. 1(b)). To determine the physical
branch, we require that $T^{(k)}(a)$ is real in the real interval
$a\in(0,a_2)$.
\begin{figure}
\unitlength=0.8mm
\begin{picture}(140,70)
\put(35,15){\vector(0,1){30}} \put(35,30){\vector(1,0){15}}
\thicklines{\put(38,38){\line(0,1){7}}}
\thicklines{\put(38,22){\line(0,-1){7}}} \put(40,38){$a_{3}$}
\put(40,22){$a_{2}$} \put(25,33){$a_{1}$}
{\linethickness{0.5mm}{\put(35,30){\line(-1,0){20}}}}
\put(35,30){\circle*{2}} \put(25,30){\circle*{2}}
\put(38,39){\circle*{2}} \put(38,21){\circle*{2}}
\put(35,47){Im($a$)} \put(50,32){Re($a$)} \put(25,10){$0<n_{f}\leq
7$} \put(15,10){(a)}
\thinlines{\put(100,15){\vector(0,1){30}}}
\linethickness{0.5mm}{\put(100,30){\line(-1,0){25}}}
\thinlines{\put(100,30){\line(1,0){8}}}
\linethickness{0.5mm}{\put(108,30){\line(1,0){20}}}
\put(100,30){\circle*{2}} \put(108,30){\circle*{2}}
\put(120,30){\circle*{2}} \put(90,30){\circle*{2}}
\put(100,47){Im($a$)} \put(125,32){Re($a$)} \put(90,10){$8\leq
n_{f}\leq 16$} \put(90,32){$a_{1}$} \put(108,32){$a_{2}$}
\put(120,32){$a_{3}$} \put(80,10){(b)}
\end{picture}
\caption{The singularity structure of the function $t=T^{(4)}(a)$.
Two different situations are shown. Branch cuts are represented by
bold lines and branch points by the blobs.}
\end{figure}
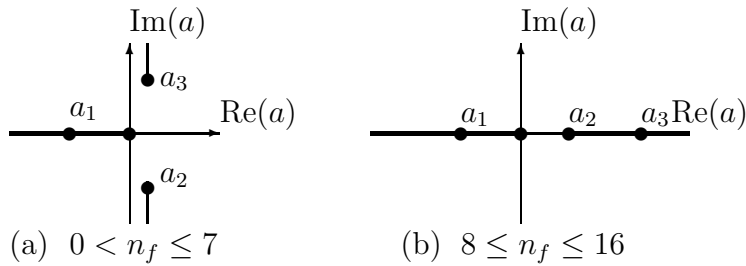
We  may now analyze the singularity structure of the running
coupling $a={\tilde a}(t)\equiv{a(u)}$ in the complex $t$-plane,
and hence in the complex $u$-plane too. Evidently, the singular
points are determined by the limiting values of the function
$T^{(k)}(a)$ as $a$ tends to infinity \footnote[4]{we assume that
the coupling does not have singular points where it takes finite
values.}. In general, the limiting values may depend on the way
along which $a$ tends to infinity. Consider, for example, the
3-loop case for $6\leq n_{\rm f}\leq 16$. We start from formula
(\ref{3lim}). Let $a$ be a point in the complex plane
$a=|a|\exp{(\imath \delta)}$, where $\delta$ is the phase of $a$.
The analytical continuation to this point gives
\begin{equation}
\label{3l}
\begin{array}{l}
\fl T^{(3)}(a)=\displaystyle{\exp{(-\imath\delta)}|a|^{-1}-0.5
b_{1}\ln(|P^{(3)}(a)||a|^{-2})
-\frac{0.5b_1^2-b_2}{\sqrt{-\Delta^{(3)}}}\ln\left|\frac{a_{2}(a-a_{1})}{-
a_{1}(a_2-a)}\right|}\\
\qquad{}+\imath(b_1\delta-0.5b_1(\delta_1+\delta_2)-(0.5b_1^{2}-b_2)(\delta_1-
\delta_2)/\sqrt{-\Delta^{(3)}}),
\end{array}
\end{equation}
where $P^{(3)}(a)=b_{2}a^2+b_{1}a+1$ ($b_{2}<0$) and $\delta_{i}$
($i=1,2$) denote the increments of the arguments of
$(\tilde{a}-a_{1})$ and $(a_{2}-\tilde{a})$ as $\tilde{a}$ goes
from $0$ to the point $a$ along a contour $\Gamma$:
$\delta_{1}=\Delta_{\Gamma} \arg{(\tilde{a}- a_{1})}$ and
$\delta_{2}=\Delta_{\Gamma}\arg(a_{2}-\tilde{a})$. We can now
calculate limiting values of $T^{(3)}(a)$ as $a$ tends to the
sides of the branch cuts along the real axis.  We find, at the
sides,
\begin{equation}
\label{uplow}
\begin{array}{llll}
\delta=0,& \delta_{1}=0,& \delta_{2}=\mp\pi,&
\mbox{if\quad Re$(a)>a_{2},$ Im$(a)=\pm\epsilon$}\\
\delta=\pm\pi,& \delta_{1}=\pm\pi,& \delta_{2}=0,& \mbox{if\quad
Re$(a)<a_{1}$, Im$(a)=\pm\epsilon$}
\end{array}
\end{equation}
where $\epsilon\rightarrow 0^{+}$ is assumed. Using
Eq.~(\ref{3l}) with Eq.~(\ref{uplow}), we may easily calculate the
limits of $T^{(3)}(a)$ when $a$ goes to infinity along the upper
or lower side of the right (left) cut. One may confirm that the
result will be the same regardless of the branch cut chosen. It
depends only on the side of the cut: $
t_{\pm}^{(3)}(n_{f})=\lim_{a\rightarrow\infty}T^{(3)}(a\pm\imath
\epsilon)=\lim_{a\rightarrow-\infty} T^{(3)}(a\pm\imath \epsilon)
$. Thus we find, for $6\leq n_f\leq 16$,
\begin{equation}
\label{3lsng} \fl t_{\pm}^{(3)}(n_{\rm
f})=-0.5b_{1}\ln|b_{2}|+\frac{b_{2}-
0.5b_{1}^2}{\sqrt{-\Delta^{(3)}}}\ln|a_{2}/a_{1}|
\pm\imath\left(0.5b_{1}+\frac{b_{2}-
0.5b_{1}^2}{\sqrt{-\Delta^{(3)}}}\right)\pi.
\end{equation}
Consider now the 4-loop case for $8\leq n_{f}\leq 16$. Using
arguments similar those employed in the 3-loop case, we find the
singular points
\begin{equation}
\label{4lsng} t_{\pm}^{(4)}(n_{\rm f})
=b_{3}^{-1}\sum_{k=1}^{3}E_{k}\ln|a_{k}|
\pm\imath(E_{2}+E_{3})\pi.
\end{equation}
Now it is  sufficient to show that the obtained limits do not
depend on the special choice of the directions  in the complex
$a$-plane. To see this, let us take the contour integral in
Eq.~(\ref{gsstev}) along a closed contour chosen as follows. Let
the contour consists of two different curves with common starting
point at zero. Let both curves lie in the upper (lower) complex
plane and are connected by the arc of a circle with centre at zero
and radius $R$. The integrand has no singularity inside the
contour, and the value of the integral is therefore zero (the
Cauchy's theorem). Consider the limit when the radius of the
circle tends to infinity. Then the integral along the arc tends to
$0$. So that the integrals along the two different curves tend to
the same limit. Thus the result stated follows.

It is important to determine whether or not the singular points
$t_{\pm}^{(k)}(n_{\rm f})$ are located inside the strip
$-\pi<\Im(t)\leq \pi$. The strip is image of the first (physical)
sheet of the complex $Q^2$ plane under $t=\ln(Q^{2}/{\Lambda}^2)$.
Depending on the value of $n_{\rm f}$ there are two cases to
consider. The first case is that the points lie inside the strip
so that the unphysical Landau singularities appear in the first sheet.
Then the running coupling is not causal, and thus perturbation theory is
incomplete: the non-perturbative contributions are required to
remove the unphysical singularities \cite{ggk,ss,gr,gg}. This case
corresponds to real-world QCD, where $n_{\rm f}\leq 6$. In the
second case, the singular points may arise beyond the strip. So
that there are not real or complex singularities on the first
sheet of the momentum squared variable, and thus perturbation
theory is consistent with causality. The singularities still may
present only on the time-like axis $Q^2<0$. The second possibility
may be realized for sufficiently large $n_{\rm f}$ values. The
value of $n_{\rm f}$ above which the causal analytical structure
of the coupling is restored can be found from the equation
\begin{equation}
\label{cc} \Im\{t^{(k)}(n_{\rm f}^{\star})\}=\pm \pi.
\end{equation}
With the 3-loop formula (\ref{3lsng}), we find the solution to
(\ref{cc}) $n_{\rm f}^{*(3)}\approx{8.460}$, and with the 4-loop
formula (\ref{4lsng}) $n_{\rm f}^{*(4)}\approx{8.455}$. Thus, the
3- and 4-loop ${\overline{\rm MS}}$ scheme results almost
coincide. We remark that the 3-loop estimation was obtained
previously in \cite{gg}. The 2-loop condition for causality of the
coupling can be found in \cite{ggk}. In our notation it reads
$b_{1}(n_{\rm f})\leq -1$, this gives for the lower boundary of
the causal region the value $n^{*(2)}_{\rm f}\approx 9.68$.

Note that for $n_{\rm f}>n_{\rm f}^{*}$ the $\beta$-function has a
positive infrared stable fixed point (see Tables 2 and 4). So that the
running coupling is trapped in the range between $0$ (the
ultraviolet fixed point) and the infrared fixed point at all
energies. The fact that QCD in perturbation theory for
sufficiently large $n_{\rm f}$ values has an infrared fixed point
has long been discussed \cite{banks}. Of particular interest is
the case when the fixed value of the coupling is sufficiently
small. Then, presumably, the theory may be reliable described
within perturbation theory for all energies including infrared
region. The corresponding interval of $n_{\rm f}$ values is called
as a conformal window \cite{appel, gg,gkarl, miran}. It is
believed that there is a phase transition in QCD with respect to
$n_{\rm f}$ inside the range of validity of asymptotic freedom
$0\leq n_{\rm f}\leq 16$. For small values of $n_{\rm f}$ below
the critical point ($n_{\rm f}<N_{f}^{\rm cr}<16$) the theory is
defined via the confining phase. Above this point, there is a
conformal window $N_{\rm f}^{\rm cr}<n_{\rm f}\leq 16$, where the
theory is defined via the non-Abelian Coulomb phase with no color
confinement and dynamical chiral symmetry breaking. One possible
way to determine the critical point is to use  the
Oehme-Zimmermann criterion for the gluon confinement, the
superconvergence rule for the transverse gluon propagator
\cite{oz}. This gives the value $N_{\rm f}^{{\rm cr}}=13N_{\rm
c}/4$ (=9.75 for $N_{\rm c}=3$ colours). Other possibility is to
apply arguments of dynamical chiral symmetry breaking \cite{
appel, miran, fgms,robw}. This gives slightly large value $N_{\rm
f}^{\rm cr}\approx 4N_{\rm c}$. Assuming the value for $N_{f}^{\rm
cr}$  as predicted by the superconvergence rule, the
authors of \cite{gg,gkarl} have given arguments  that the
perturbative running coupling inside the conformal window (and beyond
the 1-loop approximation) is always causal, i.e. $n^{*}_{\rm
f}<N_{\rm f}^{\rm cr}$. Note that the infrared fixed value of the
coupling inside the window is not large: it coincides with the
root $a_{2}$ (in the 3-loop case, for $n_f=10$, $a_2\approx
0.26$).

Consider now the cases where Eq. (\ref{algeq}) has complex roots.
In the $\overline{\rm MS}$ scheme this takes place at 3-loops if
$0\leq n_{\rm f}\leq 5$ and at 4-loops if $0\leq n_{\rm f}\leq 7$.
The corresponding cuts in the $a$-plane are chosen as shown in
Fig.~ 1(a). We first calculate the limit of $T^{(k)}(a)$ as $a$
tends to infinity along a line going to infinity in the right
half-plane $\Re(a)>\Re(a_{2})$. Evidently, the result will not
depend on the particular choice of the direction as far as the
line belongs to the right-half plane $\Re(a)>\Re(a_{2})$. Choosing
the path along the positive semi-axis and using the 3- and 4-loop
formulas (\ref{3cases}) and (\ref{smalln}), we calculate the
limits, $t^{(k)}_{\rm rhp}(n_{\rm f})=\lim_{a\rightarrow\infty}
T^{(k)}(a)$,
\begin{equation}
\label{rgh3} t^{(3)}_{{\rm rhp}}(n_{\rm f})=
-0.5b_{1}\ln(b_{2})+\frac{2b_{2}-b_{1}^2}{\sqrt{\Delta^{(3)}}}\left(\frac{\pi}{2}
-\arctan\left(\frac{b_{1}}{\sqrt{\Delta^{(3)}}}\right)\right),
\end{equation}
\begin{equation}
\begin{array}{l}
\label{rgh4} t^{(4)}_{\rm{rhp}}(n_{\rm
f})=b_{3}^{-1}\left(-2\Im(E_{2})\left(0.5\pi+
\arctan\left[\frac{\Re(a_{2})}{|\Im(a_{2})|}\right]\right)\right.\\
\left.+E_{1}\ln|a_{1}|+2\Re {(E_{2})}\ln|a_2|\right),
\end{array}
\end{equation}
here the subscript ``rhp'' shows that the limits are calculated
along the way going to infinity through the right half plane
$\Re(a)>\Re(a_{2})$. Let us now calculate the limits of
$T^{(k)}(a)$ when $a$ tends to infinity through the left half
plane $\Re(a)<\Re(a_{2})$. We may take without loss of generality
the ways along the sides of the cut running on the negative
semi-axis. The limiting values of $T^{(k)}(a)$ from above and
below the cut, $T_{\pm}^{(k)}(a)=\lim_{\epsilon\rightarrow
0^{+}}T^{(k)}(a\pm\imath \epsilon)$, may be determined by the
analytical continuation of the right hand sides of
Eqs.~(\ref{3lim}) and (\ref{4loop1}). The singularities are then
determined by the limiting values $t^{(k)}_{{\rm
lhp}\pm}=\lim_{a\rightarrow -\infty}T^{(k)}_{\pm}(a)$. This gives
\begin{equation}
\label{lft3} t^{(3)}_{{\rm lhp}\pm}(n_{\rm f})=t^{(3)}_{{\rm
rhp}}(n_{\rm f})-\pi (2b_{2}- b_{1}^2)/\sqrt{\Delta^{(3)}}
\pm\imath \pi b_{1},
\end{equation}
\begin{equation}
\label{lft4} t^{(4)}_{{\rm lhp}\pm}(n_{f})=t^{(4)}_{{\rm
rhp}}(n_{f})+2\pi \Im(E_{2})/b_{3}\pm\imath\pi(b_{1}-E_{1}/b_{3}).
\end{equation}
Equivalently, by the analytical continuation of (\ref{gsstev}) for
negative values of $a$ we obtain the useful formula
\begin{equation}
\label{usf1}  \fl t^{(k)}_{{\rm lhp}\pm}(n_{\rm
f})=-b_{1}\ln{b_1}+p.v.\int_{0}^{-\infty}g^{(k)}(a')da'
\pm\imath\pi(b_{1}+{\rm res}[g^{(k)}(a),a^{(k)}_{1}]),
\end{equation}
where $a^{(k)}_{1}$ denote the real negative root of (\ref{algeq})
(which presents only in the 4-th order case) and the integral here
is considered in the ``principal value sense" (p.v.). It is seen
from the Tables 1 and 3 that in these cases the Landau
singularities present in the first sheet of the $Q^2$-plane.
\begin{table}
\caption{The 3-loop $\overline{\rm MS}$ quantities: the complex
zeros of the $\bar{\beta}$-function $a_{1,2}$, and the singularity
locations $t^{(3)}_{{\rm rhp}}$ and $t^{(3)}_{\rm{lhp}\pm}$ in the
t-plane as a function of $n_{\rm f}$ for $0\leq n_{\rm f}\leq 5$.}
\begin{indented}
\item[]\begin{tabular}{@{}llll} \hline $ n_{\rm f}$& $a_{1,2}$ &
$t_{{\rm rhp}}^{(3)}$& $t_{{\rm lhp}\pm}^{(3)}$
\\ \hline
0& $-0.393\pm 0.882\imath$& 0.844   & $-1.539\pm 2.648\imath$\\
1& $-0.399\pm 0.892\imath$& 0.839   & $-1.506\pm 2.628\imath$\\
2& $-0.415\pm 0.916\imath$  & 0.830   & $-1.433\pm 2.578\imath$\\
3& $-0.447\pm 0.965\imath$  & 0.810   & $-1.293\pm 2.482\imath$\\
4& $-0.526\pm 1.071\imath$  & 0.766   & $-1.026\pm 2.322\imath$\\
5& $-0.819\pm 1.349\imath$  & 0.650   & $-0.423\pm 2.067\imath$\\
\hline
\end{tabular}
\end{indented}
\end{table}

\begin{table}
\caption{Same as in Table 2, but for $6\leq n_{\rm f}\leq 16$.}
\begin{indented}
\item[]\begin{tabular}{@{}llll} \hline $ n_{\rm f}$& $a_{1}$ &
$a_{2}$ & $t_{\pm}^{(3)}$
\\ \hline
6& -1.49& 7.09& $ 0.173\mp 0.077\imath$\\
7& -0.877& 1.24& $ -0.159\mp 1.05\imath$\\
8& -0.651& 0.660& $ -0.019\mp 2.36\imath$\\
9& -0.509& 0.409& $ 0.620\mp 4.26\imath$\\
10& -0.403& 0.264& $ 2.17\mp 7.21\imath$\\
12& -0.245& 0.104& $ 13.6\mp 21.3\imath$\\
14& -0.125& 0.028& $ 98.7\mp 89.9\imath$\\
16& -0.023& 0.001& $6216\mp 2852\imath$\\
\hline
\end{tabular}
\end{indented}
\end{table}

\begin{table}
\caption{The 4-loop $\overline{\rm MS}$ quantities for $0\leq
n_{\rm f}\leq 7$.}
\begin{indented}
\item[]\begin{tabular}{@{}lllll} \hline $ n_{\rm f}$& $a_{1}$&
$a_{2,3}$ & $t_{{\rm rhp}}^{(4)}$& $t_{{\rm lhp}\pm}^{(4)}$
\\ \hline
0& -0.797& $0.130\mp 0.782\imath$& 1.164& $-1.294\pm 0.961\imath$\\
1& -0.794& $0.134\mp 0.784\imath$& 1.163& $-1.284\pm 0.930\imath$\\
2& -0.793& $0.142\mp 0.792\imath$& 1.158& $-1.256\pm 0.871\imath$\\
3& -0.796& $0.159\mp 0.810\imath$& 1.145& $-1.199\pm 0.768\imath$\\
4& -0.802& $0.190\mp 0.844\imath$& 1.114& $-1.099\pm 0.593\imath$\\
5& -0.806& $0.259\mp 0.906\imath$& 1.035& $-0.940\pm 0.295\imath$\\
6& -0.795& $0.444\mp 1.012\imath$& 0.821& $-0.707\mp 0.217\imath$\\
7& -0.741& $1.124\mp 0.973\imath$& -0.047& $-0.423\mp 1.074\imath$\\
\hline
\end{tabular}
\end{indented}
\end{table}

\begin{table}
\caption{Same as in Table 4, but for $8\leq n_{\rm f}\leq 16$.}
\begin{indented}
\item[]\begin{tabular}{@{}lllll} \hline $ n_{\rm f}$& $a_{1}$&
$a_{2}$ & $a_{3}$& $t_{\pm}^{(4)}$
\\ \hline
8& -0.623& 0.699& 6.476& $-0.093\mp 2.37\imath$\\
9& -0.482& 0.426& 4.810& $ 0.522\mp 4.28\imath$\\
10& -0.362& 0.281& 1.937& $ 1.95\mp 7.32\imath$\\
12& -0.193& 0.112& 0.548& $13.1\mp 21.9\imath$\\
14& -0.085& 0.030& 0.196& $98.1\mp 91.8\imath$\\
16& -0.014& 0.001& 0.031& $6219\mp 2858\imath$\\
\hline
\end{tabular}
\end{indented}
\end{table}

\section{The Proof of the Convergence of the Series}
Inserting series (\ref{ser}) into the RG equation (\ref{eff}), we
recursively determine the coefficients
$\{c_{n}^{(k)}\}_{n=3}^{\infty}$ in terms of $c_{2}$ ($c_{1}=1$)
and the $\beta$-function coefficients. However, the coefficient
$c_{2}^{(k)}$ still remains undetermined. This reflects the
arbitrariness in the definition of the $\Lambda$-parameter. With
the conventional definition of the parameter, we find that
$c_{2}^{(k)}=0$. This follows from Eqs.~(\ref{w2}) and (\ref{ser})
if we use the asymptotic expansion for the Lambert-$W$ function
(see pp. 22-23 in \cite{lamb})
\begin{equation}
\label{wasy} \fl
W_{k}(z)=L_{1}-L_{2}+L_{2}/L_{1}+L_{2}(-2+L_{2})/2L_{1}^{2}
+L_{2}(6-9L_{2}+2L_{2}^{2})/6L_{1}^{3}+O((L_{2}/L_{1})^4)
\end{equation}
where for the branch $W_{-1}(z)$ for real negative $z$
($z\rightarrow 0^{-}$ as $Q^{2}\rightarrow\infty$) we must put
$L_{1}=\ln(-z)$ and $L_{2}=\ln(-\ln(-z))$. Several first
coefficients calculated in the 4-loop case are: $c_{1}^{(4)}=1$,
$c_{2}^{(4)}=0$, $c_{3}^{(4)}=\beta_{2}/\beta_{0}$,
$c_{4}^{(4)}=\beta_{3}/{2\beta_{0}}$,
\begin{displaymath}
c_{5}^{(4)}=\frac{5}{3}\frac{\beta_{2}^2}{\beta_{0}^2}
-\frac{\beta_{1}\beta_{3}}{6\beta_{0}^2},\qquad
c_{6}^{(4)}=-\frac{1}{12}\frac{\beta_{1}{\beta_{2}}^2}{\beta_{0}^3}
+\frac{1}{12}\frac{\beta_{3}{\beta_{1}}^2}{\beta_{0}^3}
+2\frac{\beta_{2}{\beta_{3}}}{\beta_{0}^2},\ldots.
\end{displaymath}
Inserting these values of the coefficients  into series
(\ref{ser}) and using (\ref{wasy}), one may readily reproduce the
conventional asymptotic representation for the coupling (see, for
example, \cite{btk})
\begin{equation}
\label{asycoup} \fl \alpha_{{\rm
asy}}(Q^2)=\displaystyle{\frac{1}{\beta_0
L}-\frac{\beta_1}{\beta_0^3}\frac{\ln L}{L^2}}
+\displaystyle{\frac{1}{{\beta}_{0}^{3}
L^3}\left(\frac{\beta_1^2}{\beta_0^2}(\ln^2 L-\ln
L-1)+\frac{\beta_2}{\beta_0}\right)}+O\left(\frac{\ln^3
L}{L^4}\right),
\end{equation}
where $L=\ln(Q^2/\Lambda^{2})\gg 1$.

Let us change the variable according to $Q^2\rightarrow
\theta=\beta_{0}\alpha_{s}^{(2)}(Q^2)$ and introduce the new
function $w(\theta)\equiv
w^{(k)}(\theta)=\alpha_{s}^{(k)}(Q^2)/\alpha_{s}^{(2)}(Q^2)-1$.
The RG equation (\ref{eff}) may be rewritten as
\begin{equation}
\label{rge1} \theta\frac{d w}{d \theta}=f^{(k)}(\theta,w),
\end{equation}
where
\begin{equation}
\label{rhs} f^{(k)}(\theta,w)=\frac{(w+1)^{2}}{1+b_{1}\theta}
\sum_{n=0}^{k-1}b_{n}{\theta}^{n}(1+w)^{n}-(w+1),
\end{equation}
with $b_{n}=\beta_{n}/\beta_{0}^{n+1}$. The function of two
variables $f^{(k)}(w,\theta)$ has the Taylor expansion
\begin{equation}
\label{tayl}
f^{(k)}(w,\theta)=\sum_{m,n=0}^{\infty}{\eta}^{(k)}_{m,n}w^{m}\theta^{n},
\end{equation}
with ${\eta}^{(k)}_{0,0}=0$, ${\eta}^{(k)}_{1,0}=1$ and
${\eta}^{(k)}_{0,1}=0$. In the 4-loop case, the expansion is
\begin{equation}
\label{tayl4} f^{(4)}(w,\theta)=w+b_{1}\theta
w+b_{2}\theta^2+w^{2}+(b_{3}-b_{1}b_{2})\theta^3+\ldots.
\end{equation}
We may now use  the analytical theory of differential equations
\cite{ince, fl} to investigate Eq.~(\ref{rge1}). With the initial
condition $w(0)=0$, this equation has a singularity: for $\theta=0$
and $w=0$ the ratio $f^{(k)}(\theta,w)/\theta$ is undefined.
Nevertheless,   in the special case where ${\eta}^{(k)}_{0,0}=0$,
${\eta}^{(k)}_{1,0}=1$ and ${\eta}^{(k)}_{0,1}=0$,
Eq.~(\ref{rge1})   may still have an analytic solution
satisfying the initial condition
$w(0)=0$  (see e.g. \cite{ince} and \cite{fl}).
For the sake of clarity, the following discussion is quite detailed. The expansion
(\ref{tayl}) converges in the domain $D=\{0<|w|<r_{1},
0<|\theta|<r_{2}\}$, where $r_{1}$ and $r_{2}$ are some positive
numbers chosen in the range $\{r_{1},r_{2}:r_{1}<\infty, \quad
r_{2}<1/|b_{1}|\}$. It follows then from the classical theory that
there exists a positive number $M^{(k)}$ such that
$|f^{(k)}(w,\theta)|\leq M^{(k)}$ for $(w,\theta)\subset D$, and
the coefficients ${\eta}^{(k)}_{m,n}$ satisfy the inequalities
\cite{ince,fl}
\begin{equation}
\label{ineq} |{\eta}^{(k)}_{m,n}|\leq {\xi}^{(k)}_{m,n},\quad
\mbox{where}\quad \xi^{(k)}_{m,n}=M^{(k)}r_{1}^{-m}r_{2}^{-n}.
\end{equation}
Under these conditions, we will show that there exists a regular
solution to Eq.~ (\ref{rge1})
\begin{equation}
\label{rsol} w(\theta)\equiv
w^{(k)}(\theta)=\sum_{n=2}^{\infty}\bar{c}^{(k)}_{n}\theta^{n},
\end{equation}
where $\quad\bar{c}^{(k)}_{n}=\beta_{0}^{-n}c^{(k)}_{n+1}$, with
$c^{(k)}_{n}$ being the coefficients in the original series
(\ref{ser}). We recall that according to our choice $\bar
{c}^{(k)}_{1}=\beta_{0}^{-1}c^{(k)}_{2}=0$. Inserting expansions
(\ref{tayl}) and (\ref{rsol}) into Eq.~(\ref{rge1}) we recursively
determine the coefficients $\bar{c}^{(k)}_{n}$
\begin{equation}
\label{barcf} \bar{c}^{(k)}_{2}={\eta}_{0,2}^{(k)}, \quad
2\bar{c}^{(k)}_{3}={\eta}_{1,1}^{(k)}\bar{c}^{(k)}_{2}+{\eta}_{0,3}^{(k)},\ldots
\end{equation}
Consider now the auxiliary function $\tilde{w}=\tilde{w}(\theta)$
satisfying the equation
\begin{equation}
\label{tildw} \tilde{w}=f_{1}^{(k)}(\tilde{w},\theta)
\equiv\frac{M^{(k)}}{(1-\tilde{w}/r_{1})(1-\theta/r_{2})}-
M^{(k)}\left(1+\frac{\tilde{w}}{r_{1}}+\frac{\theta}{r_{2}}\right),
\end{equation}
it has the Taylor expansion
\begin{equation}
\label{tayl2} f_{1}^{(k)}(\tilde{w},\theta)=
\sum_{m=0,n=0}{\xi}_{m,n}^{(k)}\tilde{w}^{m}\theta^{n},
\end{equation}
with the coefficients ${\xi}_{m,n}^{(k)}$ defined in (\ref{ineq}).
Equation (\ref{tildw}) has a series solution
\begin{equation}
\label{sertw} \tilde{w}(\theta)\equiv {\tilde{w}}^{(k)}(\theta)=
\sum_{n=2}^{\infty}\gamma_{n}^{(k)}\theta^{n}.
\end{equation}
Inserting the expansions (\ref{tayl2}) and (\ref{sertw}) into
Eq.~(\ref{tildw}) we find the recurrence formulas
\begin{equation}
\label{gamma} {\gamma}_{2}^{(k)}={\xi}_{0,2}^{(k)}, \quad
{\gamma}_{3}^{(k)}= {\xi}_{1,1}^{(k)}
{\gamma}_{2}^{(k)}+{\xi}_{0,3}^{(k)},\ldots.
\end{equation}
Let us compare Eqs.~(\ref{barcf}) with Eqs.~(\ref{gamma}). Making
use of Eq.~(\ref{ineq}), we obtain the inequalities
$|\bar{c}_{n}^{(k)}|<\gamma_{n}^{(k)}\quad {\rm for} \quad
n=2,3\ldots$. It follows then from the comparison test that the
series (\ref{rsol}) is absolutely convergent in the disk of
convergence of the series (\ref{sertw}). Evidently, the series
(\ref{sertw}) has a positive radius of convergence. The radius is
equal to the modulus of the nearest to the origin singularity
$\theta_{{\rm nr}}^{(k)}$ of the function
$\tilde{w}=\tilde{w}(\theta)$. The value $\tilde{w}$ can be solved
explicitly from the quadratic equation (\ref{tildw}). The
singularities of the majorant function $\tilde{w}(\theta)$ are
therefore located at the zeros of the discriminant of the
quadratic equation. Hence we find the singular points
\begin{equation}
\label{sngpoints} \fl \theta^{(k)}_{1}=r_{2},\quad
\theta^{(k)}_{2}=-r_{1}r_{2}/M^{(k)}, \quad
\theta_{3,4}^{(k)}=r_{2}(-l^{(k)}/2-1\pm\sqrt{l^{(k)2}+8l}),
\end{equation}
where $l^{(k)}=1+r_1/M^{(k)}$. To obtain the best possible
estimation, we have to maximize $|\theta_{{\rm nr}}^{(k)}|$ with
respect to $r_{1}$ and $r_{2}$. The quantity
$M^{(k)}=\max_{w,\theta}{|f^{(k)}(w,\theta)|}$ may be determined
according to the maximum modulus principle. The modulus $|f^{(k)}(w,\theta)|$
takes its maximum on the circles $|w|=r_{1}$ and $|\theta|=r_{2}$.
We find that the maximum of $|f^{(k)}(w,\theta)|$ is attained for
real positive values of $w$ and $\theta$. Furthermore,
$\theta^{(k)}_{{\rm nr}}=\theta^{(k)}_{3}$.  We choose
the values $r_{1}=0.25$ and $r_{2}=0.42$ in the 3-loop
case, while $r_{1}=0.21$ and $r_{2}=0.32$ in the 4-loop case. Using
``Maple~7'' \cite{corl}, we  determine numerically the maximal
values of $|f^{(k)}(w,\theta)|$ on these circles.
For $n_{f}=3$, we have found that $M^{(3)}\approx
0.695$ and $M^{(4)}\approx 0.596$. Computing the modulus of the
numbers (\ref{sngpoints}) and comparing them, we determine the
radii of convergence of the majorant series (\ref{sertw}):
$\tilde{\rho}^{(3)}_{maj}=|\theta_{3}^{(3)}|\approx 0.045$ and
$\tilde{\rho}^{(4)}_{maj}=|\theta_{3}^{(4)}|\approx 0.033$. Thus
we have found that the radii of convergence of the original series
(\ref{ser}) in the $\overline{\rm MS}$ scheme in the 3- and 4-loop
orders,  at $n_{f}=3$, are bounded below as
$\rho^{(3)}\geq\tilde{\rho}^{(3)}_{maj}/\beta_{0}\approx 0.06$ and
$\rho^{(4)}\geq \tilde{\rho}^{(4)}_{maj}/\beta_{0}\approx 0.05$.
As we shall see, the actual values of  $\rho^{(k)}$ are
significantly larger than the obtained lower bounds.

We remark that the above proof of the convergence of the series
holds in all MS-like (massless) renormalization schemes, since in
the proof we have not used specific values of the $\beta$-function
coefficients and the condition $c_2=0$ is common for all these
schemes.

\section{Determination of the Radius of Convergence of the Series}
By a change of variable $Q^{2}\rightarrow \theta=a^{(2)}(u)$
($u=Q^{2}/\Lambda^{2}$) Eq.~(\ref{coup1}) can be rewritten
\begin{equation}
\label{cpl1} \frac{{\rm d} a}{{\rm d}
\theta}=\frac{\sum_{n=0}^{k-1} b_{n} a^{n+2}}{\theta^2+b_{1}
\theta^3},
\end{equation}
where $a=F^{(k)}(\theta)= a^{(k)}(u)\equiv\beta_{0}\alpha_{{\rm
s}}^{(k)}(Q^2)$. In the following we will sometimes, but not
always, omit the superscript $``(k)"$ referring to the order of
perturbation theory. In the preceding section, we have shown that
the series (\ref{rsol}) or equivalently the series
\begin{equation}
\label{ser2} a=F(\theta)=\sum_{n=1}^{\infty}{\tilde
c}_{n}{\theta}^{n}, \qquad ({\tilde c}_{n}=\beta_{0}^{-n+1}c_{n})
\end{equation}
has a positive convergence radius. It is possible then to define
the inverse function $\theta=F^{-1}(a)$, which can be expanded in
powers of $a$
\begin{equation}
\label{ser3} \theta=F^{-1}(a)=\sum_{n=1}^{\infty} d_{n}{a}^{n}.
\end{equation}
By using arguments similar to those employed in Sect.~3, one can
verify that the series (\ref{ser3})  also has a finite radius of
convergence. Under this condition, we may apply the classical
method for estimating the convergence radius of series (see the
book \cite{courant} pp. 146-148). The main argument is that the
function $a=F(\theta)$ must have at least one singular point on
the circle of convergence of the series (\ref{ser2}). There are
two possible cases. First, suppose that $\theta_{0}$ be a finite
singularity of $F(\theta)$, where the function takes a finite
value, $a_{0}=F(\theta_{0})<\infty$, while its derivative does not
exist. In terms of the inverse function $\theta=F^{-1}(a)$ these
conditions read
\begin{equation}
\label{first} \left.\frac{{\rm d}F^{-1}(a)}{{\rm d}
a}\right|_{a=a_0}=0.
\end{equation}
Using Eq.~(\ref{cpl1}) at $\theta=\theta_{0}$, we may rewrite
Eq.~(\ref{first}) in the form
\begin{equation}
\label{sngs} \left.\frac{{\rm d}\theta}{{\rm d} a}\right|_{a=a_0}=
\frac{\theta_{0}^2 (1+b_1 \theta_{0})}{\sum_{0}^{k-1} b_{n}
a_{0}^{n+2}}=0,
\end{equation}
for a finite $a_{0}$ (which is not a root of $\sum_{0}^{k-1}
b_{n}a_{0}^{n}=0$) Eq.~(\ref{sngs}) has only two solutions:
$\theta_{0}=0$ and
\begin{equation}
\label{} \theta_{0}=-1/b_{1}\qquad(=-81/64 \quad \rm{for}\qquad
n_{f}=3).
\end{equation}
The solution $\theta_{0}=0$ must be rejected, since at $\theta=0$
 Eq.~(\ref{sngs}) does not hold because of the initial
condition $F(\theta)/\theta\rightarrow 1$ as $\theta\rightarrow 0$
(see (\ref{ser2})). Secondly, suppose that there exists a curve
$C$ going to infinity in the domain of analyticity of
$\theta=F^{-1}(a)$ such that $  F^{-1}(a)\rightarrow\theta_{\rm
s}<\infty$ as $a$ tends to infinity along this curve. Then,
$F(\theta)$ has a singularity at $\theta=\theta_{\rm s}$.

First, we shall consider Eq.~(\ref{cpl1}) in the 3- and 4-loop
orders for $0\leq n_{f}\leq 5$ and $0\leq n_{f}\leq 7$
respectively. Let us integrate Eq.~(\ref{cpl1}) in the real range
$\{\theta,a:\theta>0,a>0\}$. We write the result in the
symmetrical form
\begin{equation}
\label{gsol1}  1/\theta-b_{1}\ln(b_{1}+1/\theta)=1/a-
b_{1}\ln(b_{1}+1/a)+\int_{0}^{a}g^{(k)}(a')da',
\end{equation}
where the function $g^{(k)}(a)$ is defined by (\ref{gk}), and we
have determined the integration constant according to the previous
choice (see Sect.~2). Equation (\ref{gsol1}) may be continued for
complex values of $a$ and $\theta$. Then the integral with respect
to $a$ should be regarded as a line integral in the complex
$a$-plane. The contour connecting the origin to $a$ must avoid
singular points of the integrand.
\begin{figure}
\centering \resizebox{8 cm}{6cm }{\includegraphics{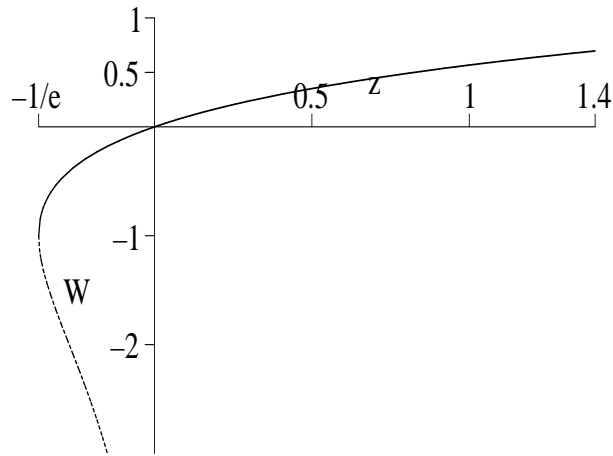}}
\caption{The two real branches of $W(z)$. The solid line
corresponds to $W_{0}(z)$ and the dotted line in the range $W<-1$
refers to $W_{-1}(z)$.}
\end{figure}

Note that all the coefficients in series (\ref{ser2}) are real, so
are the coefficients in the inverse series (\ref{ser3}).
Therefore, $\theta$ as a function of $a$ must be real for real
values of $a$ provided that $a$ is any point inside the circle of
convergence of the inverse series. It is seen from  Eq.~
(\ref{cpl1}) that there exists a real neighborhood of the origin
$\theta=0$, where the function $F(\theta)$ is real and strictly
increasing. Thus, in the 3-loop case, the derivative
$F'(\theta)>0$ if $\theta>-1/b_{1}$, provided that $0\leq
n_{f}\leq 5$ ($b_{1}>0$). In the 4-loop case, the same is true if
$\theta>\max(-1/b_{1},a_{1}^{(4)})$, where $a_{1}^{(4)}$ is the
real negative root of the equation (\ref{algeq}),
($a_{1}^{(4)}(n_{f}=3)\approx-0.796$). From this with the initial
condition $F^{-1}(0)=0$, it follows that there exists a
sufficiently small real interval including the origin, where the
function $\theta=F^{-1}(a)$ is real and strictly increasing.
Inside this interval $F^{-1}(a)>0$ ($F^{-1}(a)<0$) if $a>0$
($a<0$). Fortunately, we may solve the transcendental equation
(\ref{gsol1}) for $\theta$ explicitly as a function of $a$ in
terms of the Lambert-W function
\begin{equation}
\label{wsol1} \theta=F^{-1}(a)=-b_{1}^{-1}(1+W_{n}(z))^{-1},
\end{equation}
where $z=\zeta(a)=-(eb_{1})^{-1}\exp(-T(a)/b_{1})$, and
\begin{equation}
\label{rhs1}
T(a)=a^{-1}-b_{1}\ln(b_{1}+a^{-1})+\int_{0}^{a}g(a')\, da'.
\end{equation}
It follows from the above discussion that in the region $a>0$
inside the convergence disc of the series (\ref{ser3}) (where
$\theta>0$) the required branch in Eq.~(\ref{wsol1}) is
$W_{-1}(z)$, so that
\begin{equation}
\label{branch1}
\theta=F^{-1(k)}(a)=-b_{1}^{-1}(1+W_{-1}(\zeta^{(k)}(a)))^{-1}.
\end{equation}
Formula (\ref{branch1})  may be continued beyond the convergence
circle on the positive $a$-semi-axis. It follows from (\ref{rhs1})
that the function $z=\zeta(a)$ is negative and monotonically
decreasing in the infinite interval $a\in (0,\infty)$.  For the
considered values of $n_{f}$  (with the exception of $n_{f}=7$
case at 4-loops) $\zeta(a)$ takes  values  in the range $
(-{e}^{-1},0^{-})$ for $a\in (0,\infty)$.  The branch $W_{-1}(z)$
is real and negative with $W_{-1}(z)\in (-\infty,-1)$ for $z\in
(-{e}^{-1},0^{-})$ (see Fig.~2). Therefore, the function
$\theta=F^{-1}(a)$ determined by (\ref{branch1}) is real and
positive in the entire positive $a$-semi-axis. However, in the
4-loop case at $n_{\rm f}=7$, we find that $z\in
(-{e}^{-1},0^{-})$ only inside the interval $a\in (0,a_{{\rm b}})$
where $a_{{\rm b}}\approx 1.003$, and $z<-e^{-1}$ if $a>a_{\rm b}$.
Since $W_{-1}(z)$ has a branch point at $z=-e^{-1}$, the function
$F^{-1(4)}(a)$ at $n_{\rm f}=7$ will have a branch point at
$a=a_{\rm b}$. The corresponding branch cut may be chosen along
the positive interval $\{a: a_{\rm b}<a<\infty\}$. Formula
(\ref{branch1})  still holds on the upper side of the cut: on
the sides of the cut the function $F^{-1(4)}(a)$ at $n_{\rm
f}=7$ takes complex conjugate values. Using (\ref{branch1}), we
calculate the limit of $F^{-1}(a)$ as $a\rightarrow\infty$ along
the positive $a$-semi-axis. So we find  a singularity of the
function $a=F(\theta)$
\begin{equation}
\label{rghth} {\theta}_{{\rm s.}1}^{(k)}=
-b_{1}^{-1}(1+W_{-1}(\zeta^{(k)}(\infty))^{-1}=
a^{(2)}(u^{(k)}_{{\rm rhp}}),
\end{equation}
here we have used the formula
\begin{equation}
\label{rght1} \fl
\zeta^{(k)}(\infty)=\lim_{a\rightarrow{+\infty}}\zeta^{(k)}(a)
=-(b_{1}e)^{-1}\exp(-T^{(k)}(\infty)/b_{1})=-(b_{1}e)^{-1}(u_{\rm
rhp}^{(k)})^{-1/b_{1}},
\end{equation}
where $u^{(k)}_{{\rm rhp}}=\exp{(t^{(k)}_{{\rm rhp}})}$ being the Landau
singularity located on the positive $u$-semi-axis (see the 3-and
4-loop formulas (\ref{rgh3}) and (\ref{rgh4})). The last equality
in (\ref{rght1}) follows from Eq.~(\ref{gsstev}). In the 4-loop order, at
$n_{f}=7$, we find a pair of complex conjugate singular points,
$\theta_{{\rm s.}1\pm}=\lim_{a\rightarrow+\infty}\Theta(a\pm\imath
0)$, where $\theta_{{\rm s.}1+}$ is determined by (\ref{rghth}).

Formally we may continue Eq.~ (\ref{gsol1}) for negative real
values of the variables in the region $\{a,\theta:-1/b_{1}<a<0,
-1/b_{1}<\theta<0\}$. Assuming that each logarithm in
Eq.~(\ref{gsol1}) have its principal value, we obtain in this
region the equation
\begin{equation}
\label{gsol2}
1/\theta-b_{1}\ln(1/|\theta|-b_{1})=1/a-b_{1}\ln(1/|a|-
b_{1})+\int_{0}^{a}g(a')\, d a'.
\end{equation}
Note that the right-hand side of Eq.~(\ref{gsol2}) is in fact
regular at $a=-1/b_{1}$: the logarithmic singularities of the last
two terms are cancelled in the sum. For negative values of $a$,
the path of integration of the integral on the right of
(\ref{gsol2}) goes along negative $a$-axis, but avoids the poles
of $g(a)$ by small semi-circles above or below the axis. Equation
(\ref{gsol2}) has exactly one real negative solution for $\theta$
inside the interval $a\in (\tilde{a},0)$, where
$\tilde{a}=-\infty$ in the 3-loop order and it is the finite
negative root of (\ref{algeq}) in the 4-loop order. This solution
is determined in terms of the branch $W_{0}(z)$
\begin{equation}
\label{branch2}
\theta=F^{-1(k)}(a)=-b_{1}^{-1}(1+W_{0}(\tilde{\zeta}^{(k)}(a)))^{-1},
\end{equation}
where
$\tilde{\zeta}^{(k)}(a)=(eb_{1})^{-1}\exp(-\tilde{T}^{(k)}(a)/b_{1})$
and
\begin{equation}
\label{mphi} \tilde{T}^{(k)}(a)=a^{-1}-b_{1}\ln(-a^{-1}-b_{1})
+\int_{0}^{a}g^{(k)}(a')\,d a'.
\end{equation}
It is instructive to check that our choice for the branches on the
real $a$-axis in fact follows from the analytical continuation. To
see this, let us expand expressions (\ref{branch1}) and
(\ref{branch2}) as $a\rightarrow 0^{+}$ and $a\rightarrow 0^{-}$
respectively. We must use expansion (\ref{wasy}) for $W_{-1}(z)$
as $z\rightarrow 0^{-}$.  The same formula, but with
$L_{1}=\ln{\tilde z}$ and $L_{2}=\ln{\ln{\tilde z}}$
 holds for
$W_{0}({\tilde z})$ as $ {\tilde z}\rightarrow\infty $ (see
formula (4.19) in ref.~\cite{lamb}).
One may verify  that both expansions reproduce
the same convergent power series (\ref{ser3}). So that
Eqs.~(\ref{branch1}) and (\ref{branch2}) represent the same
analytical function in two different regions. Let us now discuss
the analytical structure of the function $\theta=F^{-1}(a)$
starting from Eq.~(\ref{wsol1}). In general, $F^{-1}(a)$ may have
singularities at the same points where $T(a)$ is singular.
Nevertheless, as we have shown, $F^{-1}(a)$ is regular at $a=0$,
while $T(a)$ is singular there. Furthermore, $F^{-1}(a)$ may have
additional singularities $a_{b\pm}$ arising due to the common
branch point of $W_{0}(z)$ and $W_{\pm 1}(z)$ at $z=-e^{-1}$. To
determine locations of these singularities we numerically solve
the equation
\begin{equation}
\label{bpeq} z=\zeta^{(k)}(a)=-e^{-1}
\end{equation}
at the 3- and 4 loop orders. The approximate locations of these
singularities for different $n_{\rm f}$ values are given in Table
5.
\begin{table}
\caption{Locations of the extra singularities in the $a$-plane at
the 3-and 4-loop orders.}
\begin{indented}
\item[]\begin{tabular}{@{}lllll}\hline $n_{\rm f}$&
$a^{(3)}_{b\pm}$& $|a^{(3)}_{b\pm}|$& $a^{(4)}_{b\pm}$
&$|a^{(4)}_{b\pm}|$
\\
\hline
0& $0.038\pm 0.732\imath$ & 0.733 & $0.156\pm 0.600\imath $&0.616\\
1& $0.038\pm 0.741\imath$ & 0.742 & $0.159\pm 0.600\imath $&0.620\\
2& $0.039\pm 0.762\imath$ & 0.763 & $0.166\pm 0.609\imath $&0.632\\
3& $0.042\pm 0.806\imath$ & 0.807 & $0.179\pm 0.629\imath $&0.654\\
4& $0.046\pm 0.902\imath$ & 0.904 & $0.207\pm 0.668\imath $&0.700\\
5& $0.060\pm 1.186\imath$ & 1.187 & $0.275\pm 0.744\imath $&0.793\\
6&  & &$0.498\pm 0.888\imath $&1.018\\
7&  & &1.003&1.003\\
\hline
\end{tabular}
\end{indented}
\end{table}
Note that not all roots of Eq.~(\ref{bpeq}) give rise to the
singularities of $F^{-1}(a)$ on the first sheet of the Riemannian
surface. Thus in the 4-loop case, for $n_{\rm f}=7$,
Eq.~(\ref{bpeq}) has the roots $a_{{\rm b}\pm}\approx 0.16\pm
0.40\imath$. But one may verify that  these points are not singular
on the first sheet. To make the function $\theta=F^{-1}(a)$
single valued, we must draw cuts in the complex $a$-plane taking
into account the branch points of $T(a)$ and those of $W$
function. We draw cuts in the complex $a$-plane attached to the
complex conjugate branch points  (say $a_{2,3}$ in the 4-loop case,
the roots of (\ref{algeq})) and running to infinity parallel to
the imaginary $a$ axis: $\{a: \Re(a)=\Re(a_{2,3}),\quad-
\infty<\Im(a)<\Im(a_{2}), \quad \Im(a_{3})<\Im(a)<\infty\}$. In
the 4-loop case, we must draw an extra cut attached to the real
branch point at $a_{1}<0$ and  running along the negative
semi-axis $\{a:-\infty<a<a_{1}\}$. We also make the branch cuts
running parallel to the imaginary axis $\{a: \Re(a)=\Re(a_{{\rm
b}\pm}), -\infty<\Im(a)<\Im(a_{{\rm b}-}), \Im(a_{{\rm}
b+})<\Im(a)<\infty\}$ attached  to the branch points at
$a_{b\pm}$, the roots of Eq.~(\ref{bpeq}). However, in the 4-loop case, for $n_{\rm f}=7$, the
branch cut must be chosen along the positive real axis
$\{a:1.003<a<\infty\}$. With this choice of the cuts, the function
$\theta=F^{-1}(a)$ will be analytic in the disc with centre the
origin and radius $r_{{\rm c}}=\min\{|a_{{\rm b}\pm}|,|a_{i}|\}$
($a_{i}$, $i=1,2\ldots$ denote the roots of Eq.~(\ref{algeq})). For
$0\leq n_{\rm f}\leq 5$, the points $a_{{\rm b}\pm}$ are closest
to the origin singularities both in  the 3- and 4-loop orders. Let
us define $a=r\exp{(\imath\delta)}$ and
$z=\zeta(a)=|z|\exp{(\imath\Phi)}$, where
\begin{equation}
 |z|=(e b_{1})^{-1}\exp(-\Re
T(a)/b_{1})\qquad\mathrm{and}\qquad \Phi=\pi-\Im T(a)/b_{1}.
\end{equation}
Let $a$ describes  the semi-circle of radius $r<r_{\rm c}$ lying
in the upper half plane ($0\leq \delta \leq \pi$) starting from
the positive semi-axis. Then the image under $z=\zeta(a)$
describes a curve in the $z$-plane. This curve intersects the real
negative $z$-semi-axis for two or more times at different points.
The number of the intersections depends on the value of $r$: it
increases when $r$ decreases. At the intersections the boundary of
the branch of $W$ is reached, so that the branch of $W$-function
must be changed when $z$ passes  these points. To define the
analytical continuation along the semi-circle, we demand that the
function $\theta=F^{-1}(a)\equiv\tilde{F}^{-1}(\zeta(a))$ will be
continuous as a function of the phase of $a$. This will be
achieved,  if we use the rules of counter-clockwise continuity
\cite{lamb} for selecting the branches  of $W$ when the curve
crosses the branch cut. These rules are given by
\begin{equation}
\label{ccc}
\begin{array}{ll}
W_{-1}(x+\imath 0)=W_{1}(x-\imath 0)& {\rm if}\quad -e^{-1}<x<0 \\
W_{1}(x-\imath 0)=W_{0}(x+\imath 0)& {\rm if}\quad -\infty<x<-e^{-1} \\
W_{n}(x+\imath 0)=W_{n+1}(x-\imath 0)& {\rm if}\quad
-\infty<x<0\quad {\rm and} \quad n\geq 1.
\end{array}
\end{equation}
We start at $a=r<r_{{\rm c}}$ on the positive semi-axis with the
branch $W_{-1}(z)$ and pass the semi-circle $\{\delta: 0\leq
\delta\leq\pi\}$ selecting the relevant branches according to the rules (\ref{ccc}). Using
``Maple 7"  \cite{corl}, we have plotted the function $z=\zeta(a)$
along the semi-circles for various values of $r$ in the interval
$0<r<r_{\rm c}$. In this way, we have determined the variations of the
phase $\Phi=\arg(z)$ along the semi-circles. Then we have  confirmed
that the analytical continuation on the
negative interval $-r_{c}<a<0$, with
the rules (\ref{ccc}), really leads to the branch $W_{0}(z)$.

Having the analytical structure of $F^{-1}(a)$ established, we can
construct explicit expressions for $F^{-1}(a)$ in the entire cut
complex $a$-plane. This enables us to calculate the limits of
$F^{-1}(a)$ as $a$ tends to infinity along different directions in
the complex plane and  determine thereby the singularities of
the function $a=F(\theta)$ in the complex $\theta$-plane.  Using
the arguments based on Cauchy's theorem  (see Sect.~2),  one may
justify that it is sufficient to calculate the limits choosing the
directions only along the real $a$-axis.

Let us now define the analytical continuation along entire
negative $a$-semi-axis. In the 3-loop case, we may represent
(\ref{mphi}) in the form
\begin{equation}
\label{f3} \tilde{T}^{(3)}(a)=a^{-1}-b_{1}\ln(||a|^{-1}-b_{1}|)+
{\rm p.v.}\int_{0}^{a}g^{(3)}(a')\,da'
\end{equation}
for all $a<0$. Thus $\tilde{T}^{(3)}(a)$ is real, and therefore
$\tilde{\zeta}^{(3)}(a)>0$ for all $a<0$. Hence
$W_{0}(\tilde{\zeta}^{(3)}(a))>0$ (see Fig.~2), so that
$F^{-1(3)}(a)<0$ for all $a<0$. It is evident  that the
required branch, in this case, will be $W_{0}(z)$ on the entire negative
$a$-semi-axis. Making $a\rightarrow-\infty$ in Eq.~(\ref{branch2}) and
using Eq.~(\ref{f3}), we find the real singular point
\begin{equation}
\label{sp-3} \theta^{(3)}_{{\rm s.}2}=\lim_{a\rightarrow
-\infty}F^{-1(3)}(a)=-(b_{1}(1+W_{0}(\tilde{\zeta}^{(3)}(-\infty)))^{-1}
\end{equation}
where $\tilde{\zeta}^{(3)}(-\infty)=(e
b_{1})^{-1}\exp(-\tilde{T}^{(3)}(-\infty)/b_{1}),$
and
\begin{equation}
\label{f-3} \fl {\tilde T}^{(3)}(-\infty)=\lim_{a\rightarrow
-\infty}\tilde{T}^{(3)}(a)= -b_{1}\ln{b_{1}}- {\rm
p.v.}\int_{-\infty}^{0}g^{(3)}(a)\,d a=\Re(t^{(3)}_{{\rm
lhp.}\pm}),
\end{equation}
the last equality in Eq.~(\ref{f-3}) follows from
Eq.~(\ref{usf1}).

In the 4-loop order, the function $\tilde{T}^{(4)}(a)$ is real
only inside the finite interval $(a_{1},0)$ of the negative
semi-axis, where $a_{1}$ is the root of (\ref{algeq})
($a_{1}\approx -0.796$ for $n_{f}=3$). The function has a logarithmic branch
point at $a=a_{1}$. The corresponding branch cut may be chosen
along the infinite interval $(-\infty,a_{1})$. To continue
$\tilde{T}^{(4)}(a)$  in the complex $a$ plane, we use  formula
(\ref{mphi}).
\begin{table}
\caption{Positions of the singularities in the $\theta$-plane at
the 3-loop order.}
\begin{indented}
\item[]\begin{tabular}{@{}lllllll}\hline $n_{f}$& 0& 1& 2& 3& 4& 5
\\ \hline
$\theta_{0}$& -1.186& -1.195& -1.219& -1.266& -1.353& -1.520\\

$\theta_{{\rm s.}1}^{(3)}$&0.627 &0.635 &0.653 &0.691 &0.776
&1.029 \\
$\theta_{{\rm s.}2}^{(3)}$&-0.594 &-0.601 &-0.618 &-0.653 &-0.731 &-0.956 \\

${\tilde\rho}$&0.594 &0.601 &0.618 &0.653 &0.731 &0.956 \\
\hline
\end{tabular}
\end{indented}
\end{table}
The limiting values of this analytic function from above and below
the left-hand cut are given by
\begin{equation}
\label{f4} \tilde{T}^{(4)}_{\pm}(a)=1/a-b_{1}\ln(|b_{1}-1/|a||)
+{\rm p.v.}\int_{0}^{a}g^{(4)}(s){\,d}s\pm \imath \kappa \pi,
\end{equation}
where $\kappa$ stands for the residue
\begin{displaymath}
\fl \kappa=\lim_{a\rightarrow a_{1}}
(a-a_{1})g^{(4)}(a)=(b_{2}+b_{3}a_{1})
\{b_{3}(1+b_{1}a_{1})(a_{1}-a_{2})(a_{1}-a_{3})\}^{-1},
\end{displaymath}
and $a_{i}$, i=1..3, denotes the roots of (\ref{algeq}). Formula
(\ref{branch2}) enables us to define the analytical continuation
of the function $\theta=F^{-1(4)}(a)$ in the cut complex $a$
plane. In particular, we need to calculate the boundary values of
$F^{-1(4)}(a)$ as $\rm{Im}(a)\rightarrow 0^{\pm}$ along the
left-hand cut $\{a: -\infty<a<a_{1}\}$. It is easy to convince
that the required branch of $W$ along the sides of the left-hand
cut will be $W_{0}(z)$, provided that $|\kappa/b_{1}|<1$. This
condition holds only for $0\leq n_{f}\leq 5$ (for example,
$\kappa/b_{1}\approx 0.690$ at $n_{f}=3$). Therefore, for $0\leq
n_{f}\leq 5$, formula (\ref{branch2}) is valid also along the
sides of the cut. But, for $6\leq n_{f}\leq 7$, we have
$|\kappa/b_{1}|>1$.  Then one may check  that the
relevant branches on the opposite sides of the cut should be $W_{\pm
1}$. We may now calculate the limit of $F^{-1(4)}(a)$ as $a$
approaches infinity going along the upper or lower side of the  cut. So we
determine the singularities of $F^{(4)}(\theta)$. We use
Eq.~(\ref{branch2}) with Eq.~(\ref{f4}) if $0\leq n_{f}\leq 5$.
However, for $6\leq n_{f}\leq 7$, the branch $W_{0}$ in
Eq.~(\ref{branch2}) should be replaced by $W_{\pm 1}$. Then we find
the singular points
\begin{equation}
\label{sp4} \fl {\theta}^{(4)}_{{\rm s.}2\pm}= \lim_{a
\rightarrow-\infty}{F}^{-1(4)}(a\pm\imath 0)=
\left\{\begin{array}{ll}
-b_{1}^{-1}(1+W_{0}(\tilde{\zeta}_{\pm}^{(4)}(-\infty)))^{-1}
&\mbox{if $0\leq n_{\rm f}\leq 5$}\\
-b_{1}^{-1}(1+W_{\pm 1}(\tilde{\zeta}_{\pm}^{(4)}(-\infty)))^{-1}
&\mbox{if $6\leq n_{\rm f}\leq 7$}
\end{array}
\right.
\end{equation}
where ${\tilde{\zeta}}_{\pm}^{(4)}(-\infty)= (e
b_{1})^{-1}\exp(-\tilde{T}^{(4)}_{\pm}(-\infty)/b_{1})$, and by
Eqs.~(\ref{f4}) and (\ref{usf1})
\begin{equation}
\label{f-4}
\tilde{T}^{(4)}_{\pm}(-\infty)=\lim_{a\rightarrow-\infty}
\tilde{T}^{(4)}(a\pm\imath 0)= \Re(t_{{\rm lhp}\pm})\pm\imath
\pi\kappa,
\end{equation}
here the subscript ``$\pm$" shows that the limits were evaluated
keeping the upper (lower) side of the cut. Evidently,
$\theta^{(4)}_{{\rm s.2}-}=\overline{\theta}^{(4)}_{{\rm s.}2+}$.

In Tables 6 and 7, we tabulate the singularities
$\theta_{0}(n_{\rm f})=-b_{1}^{-1}$, $\theta_{{\rm
s.}1}^{(k)}(n_{\rm f})$ and $\theta_{{\rm s.}2}^{(k)}(n_{\rm f})$
at the 3- and 4-loop orders respectively. In the 3-th order, we see that the
singular points $\theta_{{\rm s.}2}^{(3)}$ are closer to the
origin than the points $\theta_{0}$ or $\theta_{{\rm s.}1}^{(3)}$,
so that the radius of convergence of the series (\ref{ser2}) is
equal to $|\theta_{{\rm s.}2}^{(3)}|$. On the contrary, in the 4th
order, we find that $\tilde{\rho}^{(4)}=|\theta_{{\rm
s.}1}^{(4)}|$ if $0\leq n_{\rm f}\leq 5$, and
$\tilde{\rho}^{(4)}=|\theta_{{\rm s.}2}^{(4)}|$ if $6\leq n_{\rm
f}\leq 7$. We recall that the radius of convergence of the
original series (\ref{ser}) is determined by
${\rho}^{(k)}=\tilde{\rho}^{(k)}/\beta_{0}$ (${\rho}^{(3)}=0.965$
and ${\rho}^{(4)}=0.720$ for $n_{\rm f}=3$).

Looking at the numbers in Tables 6 and 7 one sees that the
convergence radii in the range $0\leq n_{\rm f}\leq 6$ increases
as $n_{f}$ increases. But, for fixed values of $n_f$, they
decrease as the order of perturbation theory increases.
\begin{table}
\caption{Positions of the singularities in the $\theta$-plane at
4-loops. $n_{1}$ and $n_{2}$ denote the labels of the branches of
the $W$ function used to calculate  the singularities
$\theta_{{\rm s.}1}$ and $\theta_{{\rm s.}2\pm}$ respectively.
$\tilde{\rho}$ is the radius of convergence of the series
(\ref{ser2}).}
\begin{indented}
\item[]\begin{tabular}{@{}llllllllll}\hline $n_{\rm f}$&
$\theta_{0}$& $\theta_{{\rm s.}1}$& $|\theta_{{\rm s.}1}|$&
$n_{1}$&&
$\theta_{{\rm s.}2\pm}$& $|\theta_{{\rm s.}2\pm}|$& $n_{2}$&$\tilde{\rho}$ \\
\hline
0&-1.186& 0.485&-&-1&&$-0.545\mp 0.334\imath$& 0.639& 0& 0.485\\
1&-1.195& 0.488&-&-1&&$-0.544\mp 0.341\imath$& 0.642& 0& 0.488\\
2&-1.219& 0.497&-&-1&&$-0.546\mp 0.354\imath$& 0.650& 0& 0.497\\
3&-1.266& 0.516&-&-1&&$-0.550\mp 0.380\imath$& 0.668 & 0& 0.516\\
4&-1.353& 0.554&-&-1&&$-0.552\mp 0.429\imath$& 0.699 & 0& 0.554\\
5&-1.520& 0.641&-&-1&&$-0.533\mp 0.526\imath$& 0.748& 0&0.641\\
6&-1.885& 0.934&-&-1&&$-0.394\pm 0.672\imath$& 0.779& $\pm 1$& 0.779\\
7&-3.008& $-0.887\mp 1.531\imath$&1.769& $\mp 1$&& $-0.105\pm
0.614\imath$&0.623&$\pm
1$&0.623\\
\hline
\end{tabular}
\end{indented}
\end{table}
In order to examine the obtained formulas, we calculate
numerically  the coefficients of the series (\ref{ser2}) $\tilde
{c}_{n}$ for large values of $n$. In Table 8, we study the
behaviour of the quantity ${\tilde{\rho}}_{n}^{(k)}=({|\tilde
{c}|}_{n}^{(k)})^{-1/n}$ in the 3- and 4-loop orders at $n_{f}=3$.
It is seen from the table that our predictions are in  good
agreement with the expected limiting relation
${\tilde{\rho}}_{n}^{(k)}\rightarrow {\tilde\rho}^{(k)}$ as
$n\rightarrow\infty$. The theoretical predictions are
${\tilde\rho}^{(3)}=0.653$ and ${\tilde\rho}^{(4)}=0.516$ for
$n_{\rm f}=3$ (see Tables 6 and 7).
\begin{table}
\caption{Numerical examination of the series (\ref{ser2}) to 3-th
and 4-th orders at $n_{\rm f}=3$.}
\begin{indented}
\item[]\begin{tabular}{@{}lllllll} \hline $n$&
$\tilde{\rho}^{(3)}_{n}$ & $\tilde{\rho}^{(4)}_{n}$& & $n$&
$\tilde{\rho}^{(3)}_{n}$ & $\tilde{\rho}^{(4)}_{n}$
\\ \hline
10& 1.04  & 0.737 & & 80& 0.692  & 0.550 \\
20& 0.807 & 0.632 & & 90& 0.688  & 0.547 \\
30& 0.752 & 0.597 & & 100& 0.685 & 0.544 \\
40& 0.727 & 0.579 & & 110& 0.682 & 0.542 \\
50& 0.713 & 0.568 & & 120& 0.680 & 0.540 \\
60& 0.703 & 0.560 & & 130& 0.678 & 0.538 \\
70& 0.697 & 0.554 & & 140& 0.677 & 0.537 \\
\hline
\end{tabular}
\end{indented}
\end{table}

Next consider the theoretical cases with large $n_{\rm f}$ values  where the
$\beta$-function has non-trivial real zeros. This takes place in
the 3-loop case for $n_{\rm f}=\{6-16\}$. From now on we shall
confine ourselves to the 3-loop case. There are now two different
cases which should  be considered separately. For $n_{\rm
f}=\{6-8\}$ ($b_1>0$ and $b_2<0$) the equation
$\zeta^{(3)}(a)=-e^{-1}$ has one real positive root $a_b$
($0<a_b<a_2$) and a pair of complex conjugate roots $a_{{\rm
b}\pm}$ with $\Re(a_{{\rm b}\pm})<0$ (see Table 9 ). On the real
interval $a_1<a<a_{{\rm b}}$, the real analytic solution to
(\ref{gsol1}) is
\begin{equation}
\label{l678} \theta= \left\{\begin{array}{ll}
-b_{1}^{-1}(1+W_{-1}(z))^{-1}& \mbox{if $0<a<a_{\rm b}$}\\
-b_1^{-1}(1+W_{0}(z))^{-1}& \mbox{if $a_1<a<0$,}
\end{array}
\right.
\end{equation}
where  $z={\zeta}^{(3)}(a)$
(see Eq.~(\ref{rhs1})). To determine uniquely the analytical continuation of the
right-hand side of Eq.~(\ref{l678}), we make branch cuts
on the $a$-plane. There are the branch points at $a_{1,2}$, the
roots of (\ref{algeq}), and at $a_{{\rm b}}$ and $a_{{\rm b}\pm}$,
the roots of $\zeta^{(3)}(a)=-e^{-1}$. They are listed in Table
9 as a function of $n_{\rm f}$.
\begin{table}
\caption{The location of the branch points $a_{1,2}$,  $a_{{\rm
b}}$ and $a_{{\rm b\pm}}$ of $\theta=F^{-1(3)}(a)$ for $n_{\rm
f}=\{6-16\}$.}
\begin{indented}
\item[]\begin{tabular}{@{}lllll}\hline
$n_{\rm f}$&$a_1$& $a_2$& $a_{{\rm b}}$& $a_{{\rm b\pm}}$ \\
\hline
 6&-1.489 &7.089&2.418&$-2.258\pm 0.296\imath$\\
 7&-0.877&1.238&0.813&$-0.846\pm 0.101\imath$\\
8&-0.651&0.660&0.541&$-0.552\pm 0.006\imath$\\
9 &-0.509 &0.409&-0.360& $0.439\pm 0.036\imath$ \\
10&-0.403 &0.264 &-0.253& $0.266\pm 0.034\imath$ \\
11&-0.318 &0.169 &-0.177& $0.159\pm 0.027\imath$ \\
12&-0.245 &0.104 &-0.121 &$0.089\pm 0.018\imath$ \\
13&-0.182 &0.059 &-0.078 &$0.043\pm 0.011\imath$ \\
14&-0.125 &0.028 &-0.044 &$0.015\pm 0.006\imath$ \\
15&-0.072 &0.010 &-0.020 &$0.002\pm 0.002\imath$ \\
16&-0.023 &0.001 &-0.004 &$-0.0013\pm 0.0002\imath$ \\
\hline
\end{tabular}
\end{indented}
\end{table}
We make branch cuts along the infinite intervals of the real axis
$\{a:-\infty<a<a_1\}$ and $\{a: a_{{\rm b}}<a<\infty\}$. There is
a double branch cut along the interval $\{a: a_{2}<a<\infty\}$. We
also make branch cuts along the straight lines joining the points
$a_{\rm b\pm}$ with the point $a_{1}$. Now we may continue
analytically the function (\ref{l678}) and determine its boundary
values along the sides of the right hand cut using the rules of
counter-clockwise continuity (\ref{ccc}). Choosing the ways
along the sides of this cut, we  take the limit
$a\rightarrow\infty$. Thus we find a pair of complex conjugate
singular points
\begin{equation}
\label{thsp3}
 \theta_{{\rm s.1}\pm}=\lim_{a\rightarrow\infty}F^{-1}(a\pm\imath 0)=-b_{1}^{-1}(1+W_{\pm
n}(z_{\pm}))^{-1},
\end{equation}
where $z_{\pm}=\lim_{a\rightarrow\infty}\zeta^{(3)}(a\pm\imath
0)$,
\begin{displaymath}
z_{\pm}=(eb_{1})^{-1}\exp(M_{1}\ln(|a_{1}|)/b_{1}-M_{2}\ln(a_{2})/b_{1})\exp(\mp\imath\pi
M_{1}/b_{1}),
\end{displaymath}
with $M_{1,2}=(a_{1}-a_{2})^{-1}(1+b_{1}a_{1,2})^{-1}$. It is
obvious from Cauchy's theorem that if we calculate the limits
choosing the directions along the left-hand cut, we shall reproduce
the same values, i.e. $\theta_{{\rm s.2}\pm}=\theta_{{\rm
s.1}\pm}$. Comparing the modulus of the numbers
$\theta_{0}=-b_{1}^{-1}$ and $\theta_{{\rm s.1}\pm}$ (see Table
10), we determine the radii of convergence of the series
(\ref{ser2}) for $n_{\rm f}=\{6-8\}$. In this Table, we also
indicate the required branch of the $W$ function which is used in
formula (\ref{thsp3}).
\begin{table}
\caption{The locations of the singularities of $a=F(\theta)$ and
the convergence radii of the series (\ref{ser2}) at 3-loops in the theoretical cases
when $n_{\rm f}=\{6-16\}$.}
\begin{indented}
\item[]\begin{tabular}{@{}lllll}\hline
$n_{\rm f}$ & $W_{\pm n}$ & $\theta_{0}$ & $|\theta_{{\rm s.1}\pm}|$ & $\tilde\rho$ \\
\hline
 6& $W_{\pm 1}$& 1.885& 2.114& 1.885\\
7& $W_{\pm 1}$& 3.008& 0.664& 0.664 \\
8& $W_{\pm 19}$& 48.17& 0.417&0.417 \\
9& $W_{\mp 1}$ & 2.08&0.291 &0.291 \\
10& $W_{\mp 1}$&0.761&0.208 &0.208\\
11& $W_{\mp 1}$&0.360 &0.148  &0.148 \\
12& $W_{\mp 1}$&0.180 &0.102 &0.102\\
13& $W_{\mp 1}$&0.087 &0.067 &0.067 \\
14& $W_{\mp 1}$& 0.037&0.039 &0.037\\
15& $W_{\mp 1}$&0.011 &0.018 &0.011 \\
16& $W_{\mp 1}$&0.001 &0.003&0.001\\
\hline
\end{tabular}
\end{indented}
\end{table}

Next consider the cases with  $n_{\rm f}=\{9-16\}$ ($b_1<0$,
$b_2<0$). Then the equation $\zeta^{(3)}(a)=-{e}^{-1}$ has a real
negative root $a_{\rm b}$ ($a_1<a_{\rm b}<0$) and a pair of
complex conjugate roots $a_{\rm b\pm}$ (see Table 9). The
solution to (\ref{gsol1}) which takes real values inside the real
interval $a_b<a<a_2$  is then given by
\begin{equation}
\theta=\left\{\begin{array}{ll}
-b_{1}^{-1}(1+W_{0}(z))^{-1}& \mbox{if $0<a<a_2$}\\
-b_1^{-1}(1+W_{-1}(z))^{-1}& \mbox{if $a_b<a<0$,}
\end{array}
\right.
\end{equation}
where $z=\zeta^{(3)}(a)=(e|b_1|)^{-1}\exp(T(a)/|b_1|)$. Now we
choose the cuts  along the real axis $\{a:-\infty<a<a_1\}$, $\{a:-\infty<a<a_b\}$ and
$\{a:a_2<a<\infty\}$. We also make cuts along
the straight lines joining the branch points $a_{b\pm}$ with
$a_2$. By means of the same procedure that was used  in the
previous case we calculate the locations of the singularities
$\theta_{s.1\pm}=\theta_{s.2\pm}$. They are determined by the same
formula (\ref{thsp3}). The relevant branches of $W$ function are
listed in Table 10. In this Table, we tabulated the magnitudes of
$\theta_0$, $|\theta_{s.1\pm}|=|\theta_{s.2\pm}|$ and
$\tilde\rho$, the convergence radius of the series (\ref{ser2}),
for $n_{\rm f}=\{6-16\}$.

\section{The Momentum Scale Associated With the Convergence Radius of the Series}

The convergence region of the series (\ref{ser2}) in the real momentum
squared space may be easily determined, since the mapping
$Q^{2}\rightarrow \theta=a^{(2)}(Q^2)$ for real positive
$Q^2>Q_{L}^{2}\geq 0$ is monotonic ($Q_{L}^{2}$ being the real
Landau singularity of the 2-loop coupling which presents if $0\leq
n_{\rm f}\leq 8.05$). First, we consider the series for large (mainly unphysical)
$n_{\rm f}$ values. Note that the quantity $\theta_0=-b_{1}^{-1}$
in the Banks-Zaks domain ($n_{\rm f}>8.05$) is the infrared fixed
point of the 2-loop coupling $\theta=a^{(2)}(u)$, so that
$0<\theta<|b_{1}|^{-1}$ for all $Q^2\in(0,\infty)$. From the Table
10, we see that $\tilde\rho=\theta_0$ inside the interval $n_{\rm
f}=\{14-16\}$. This means that the series (\ref{ser2}) at 3-loops
for $n_{\rm f}=\{14-16\}$ converges in the whole interval
$Q^2\in(0,\infty)$. Let $n_{\rm f}^{**}$ be the lowest value of $n_f$ for
which this equality holds ($n_{\rm f}^{**}=14$ in the
${\overline{\rm MS}}$ scheme). For $n_{\rm f}<n_{\rm f}^{**}$, the
series (\ref{ser2}) converges in the more restricted domain
$Q^{2}_{\rm min}<Q^2<\infty$ ($Q^{2}_{\rm min}>0$). The value of
$Q^{2}_{\rm min}$ may be determined from the equation
\begin{equation}
\label{Qmin}
 \theta=a^{(2)}(u)=-b_{1}^{-1}(1+W_{n}(z_{Q}))^{-1}=\tilde\rho
\end{equation}
where $z_{Q}=-(eb_1)^{-1}u^{-1/b_1}$ and $u=Q^2/{\Lambda}^2$ (see
Eq.~(\ref{w2})). Solving (\ref{Qmin}), we obtain
\begin{displaymath}
u_{\rm min}=Q^2_{\rm min}/{\Lambda}^2=
(b_1+{\tilde\rho}^{-1})^{-b_{1}}\exp({\tilde\rho}^{-1}).
\end{displaymath}
The results for the dimensionless quantity $\sqrt{u_{\rm
min}}=Q_{\rm min}/\Lambda$ $(Q_{\rm min}=\sqrt{Q_{\rm min}^{2}})$
to the 3- and 4-loop orders, for $n_{\rm f}=\{0-6\}$, are tabulated
in Table 11. We compare $\sqrt{u_{\rm min}}$ with the ratio
$\sqrt{u_{\rm rhp}}=Q_{\rm rhp}/\Lambda$, where
$Q_{\rm rhp}=\sqrt{Q_{\rm rhp}^{2}}$, and $Q_{{\rm rhp}}^2$ is the real space-like  Landau
singularity of the coupling.
\begin{table}
\caption{The  ratios $\sqrt{u_{\rm min}^{(k)}}=Q^{(k)}_{\rm
min}/\Lambda$ and $\sqrt{u_{\rm rhp}^{(k)}}=Q^{(k)}_{\rm
rhp}/\Lambda$ in the $\overline{\rm MS}$ scheme to the 3-and
4-loop orders for $n_{\rm f}=\{0-6\}$.}
\begin{indented}
\item[]\begin{tabular}{@{}lllll}\hline $n_{\rm f}$ &
${\sqrt{u_{min}^{(3)}}}$ &${\sqrt{u_{rhp}^{(3)}}}$
&${\sqrt{u_{min}^{(4)}}}$
& ${\sqrt{u_{rhp}^{(4)}}}$ \\
\hline
0& 1.571&1.525&1.790&1.790 \\
1&1.566 &1.521&1.788&1.788 \\
2&1.558 &1.514&1.784&1.784 \\
3&1.541 &1.500&1.773&1.773 \\
4&1.505 &1.467&1.745&1.745 \\
5&1.416 &1.384&1.678&1.678 \\
6&1.283 &--&1.623&1.507 \\
 \hline
\end{tabular}
\end{indented}
\end{table}
It is seen from the Table that in general  the quantity  $Q_{{\rm
min}}^{2}$ can not be identified with the real Landau singularity
$Q_{\rm rhp}^{2}$. The equality $Q_{{\rm min}}^{2}=Q_{\rm
rhp}^{2}$ occurs only in the cases where the convergence radius
$\tilde\rho$ is determined via the real (space-like) Landau
singularity.
This happens, for example, in
the $\overline{\rm MS}$ scheme in the 4-loop case for $n_{\rm
f}=\{0-5\}$. However, in the case where $\tilde\rho$ is
determined via the complex Landau singularities, $Q_{\rm
lhp\pm}^{2}$,  the relation between $Q^{2}_{{\rm min}}$ and
$Q^{2}_{\rm{lhp}\pm}$ is more complicated. Then we observe the inequality $Q^{2}_{{\rm
min}}>Q^{2}_{{\rm rhp}}$. Such a situation occurs, for instance,
in the $\overline{\rm MS}$ scheme in the 3-loop case for $n_{\rm
f}=\{0-5\}$.

It is reasonable to compare $Q_{\rm min}$ with the infrared
boundary of QCD, the momentum scale $\mu_{\rm c}$ that separates
the perturbative and nonperturbative regimes of the theory in the
confining phase.  Several estimates for this quantity was
suggested using different nonperturbative methods. In recent work
\cite{AA2} an useful nonperturbative approximation for the QCD
$\beta$-function has been constructed.  The model gives a number
consistent results for various nonperturbative quantities.
In particular, it was obtained that
$(\mu_{\rm c}/\Lambda_{\rm
QCD})_{3-loop}\approx 3.204$ and $(\mu_{\rm c}/\Lambda_{\rm
QCD})_{4-loop}\approx 3.526 $
with the perturbative ${\overline{\rm MS}}$ scheme component of the
total $\beta$-function in the 3-th and 4-th orders.
Other approach to determine the
infrared boundary is to use arguments based on dynamical chiral
symmetry breaking in QCD.
There are the results obtained within
the nonperturbative framework of Schwinger-Dyson equations
\cite{fgms,robw}. It was found in ref.~\cite{fgms} that the critical value of
the coupling needed to generate the chiral condensate is
$\alpha_{c}=\pi/4$ (for $N_{\rm c}=3$ QCD). It is reasonable to
identify the corresponding scale with the infrared boundary
\cite{cems}.  To obtain approximations to $\mu_{c}$, we may use
the perturbative expressions for the coupling in the
${\overline {\rm MS}}$ scheme. With this simplifying assumption, the equation
$\alpha_{s}^{(k)}(\mu_{c}^2)=\pi/4$ in the 3- and 4-loop orders, for $n_{\rm f}=3$,
yields the estimates
\begin{displaymath}
(\mu_{\rm c}/\Lambda_{{\overline {\rm MS}}})_{{\rm 3-loop}}=1.972
\quad {\rm and}\quad (\mu_{\rm c}/\Lambda_{{\overline {\rm
MS}}})_{{\rm 4-loop}}=2.115.
\end{displaymath}
The two different  estimates considered above are consistent with the inequality
$Q_{\rm m}^{2}<\mu_{\rm c}^{2}$.
Thus it seems  reasonable  to believe
that the series expansion (\ref{ser2}) in the $\overline{{\rm
MS}}$ scheme may be safely used  in the whole perturbative region
$\mu_{\rm c}^{2}<Q^2<\infty$.

\section{Application to Analytic Perturbation Theory}
In the Analytic Perturbation Theory (APT) approach of Shirkov and
Solovtsov, Euclidean and Minkowskian QCD observables (which depend
on the single scale)  are represented by asymptotic expansions
over non-power sets of specific functions $\{{\cal
A}_{n}^{(k)}(u)\}_{n=1}^{\infty}$ and $\{{\mathfrak
A}_{n}^{(k)}(\bar s)\}_{n=1}^{\infty}$ respectively (see
refs.~\cite{shirk,shirk2}), here $u=Q^{2}/{\Lambda}^{2}$ and $\bar
s=s/\Lambda^2$. These sets are constructed via the integral
representations  in the following way
\begin{equation}
\label{spectrl} {\cal
A}_{n}^{(k)}(u)=\frac{1}{\pi}\int_{0}^{\infty}\frac{\varrho^{(k)}_{n}
(\varsigma)d\varsigma}{\varsigma+u},\quad {\mathfrak
A}_{n}^{(k)}({\bar s})=\frac{1}{\pi}\int_{\bar
s}^{\infty}{\varrho^{(k)}_{n} (\varsigma)\over
\varsigma}d\varsigma,
\end{equation}
where the spectral densities to the $k$-th order are determined from
powers of the running coupling: $\varrho^{(k)}_{n}
(\varsigma)=-{\Im}(a^{(k)n}(-\varsigma+\imath 0))$. In APT the
power series (\ref{ser2}) give rise to the following  series of
functions
\begin{eqnarray}
\label{acalser} {\cal A}^{(k)}_{m}(u)=\sum_{n=m}^{\infty}{\cal
C}_{m,n}^{(k)}{\cal
A}_{n}^{(2)}(u)\quad m=1,2\ldots\\
\label{mathser} {\mathfrak A}_{m}^{(k)}({\bar s})=
\sum_{n=m}^{\infty}{\cal
C}_{m,n}^{(k)}{\mathfrak A}_{n}^{(2)}(\bar s)\quad m=1,2\ldots\\
\label{rhoser} \varrho^{(k)}_{m}
(\varsigma)=\sum_{n=m}^{\infty}{\cal
C}_{m,n}^{(k)}\varrho^{(2)}_{n} (\varsigma)\quad m=1,2\ldots,
\end{eqnarray}
where ${\cal C}_{m,m}^{(k)}=1$. The sets of coefficients $\{{\cal
C}_{m,n}^{(k)}\}_{n=m}^{\infty}$, $m=1,2\ldots$, are constructed
from the set of coefficients of the original series, $\{{\tilde
c}_{n}^{(k)}\}_{n=1}^{\infty}$, according to the rules for
products of power series: ${\cal C}_{1,n}^{(k)}={\tilde
c}^{(k)}_{n}$, ${\cal C}_{2,n}^{(k)}=\sum_{j=1}^{n-1}{\tilde
c}^{(k)}_{n-j}{\tilde c}^{(k)}_{j}$ etc. The spectral densities at
the 2-loop order can be expressed analytically in closed form \cite{join,mpr}
\begin{equation}
\label{clofsf} \fl {\varrho}_{n}^{(2)}(\varsigma)=
b_{1}^{-n}\Im(1+W_{-1}(z_{\varsigma}))^{-n} \quad {\rm with}\quad
z_{\varsigma}=
(eb_{1})^{-1}{\varsigma}^{-1/b_{1}}\exp{(-\imath\pi(1/b_{1}-1))}.
\end{equation}
Now we are going to prove that the series of functions
(\ref{acalser}), (\ref{mathser}) and (\ref{rhoser}) are uniformly
convergent over whole ranges of the corresponding variables:
$0<u<\infty$, $0<{\bar s}<\infty$ and $0<\varsigma<\infty$.
Suppose  that the series (\ref{rhoser}) is uniformly convergent.
Then the series (\ref{acalser}) and (\ref{mathser}) will also  be
uniformly convergent. To see this, let us insert the series
(\ref{rhoser}) into integral representations given in
(\ref{spectrl}) and integrate term-by-term. This yields the series
(\ref{acalser}) and (\ref{mathser}), which must be uniformly
convergent, as results of term-by-term integration of the
uniformly convergent series. Evidently, the factors
$1/(\varsigma+u)$ and $1/\varsigma$ inside the integrals  will not
destroy this statement.

Let us now write $W_{-1}(z_{\varsigma})={\cal W}={\cal
X}+\imath{\cal Y}$, $(1+{\cal W})^{-1}={\cal
R}\exp{(\imath\Psi)}$, where ${\cal R}=(({\cal X}+1)^2+{\cal
Y}^2)^{-1/2}$ and $\Psi=\arcsin(-{\cal Y}{\cal R})$ (for the
branch  $W_{-1}$, we have $-3\pi<{\cal Y}<0$). According to this,
we may rewrite the 2-loop spectral densities (\ref{clofsf}) as
\begin{equation}
\label{clofsf2}
 {\varrho}_{n}^{(2)}(\varsigma)=({\cal
R}/b_{1})^{n}\sin(n\Psi), \quad n=1,2\ldots.
\end{equation}
It is seen from Eq.~(\ref{clofsf2}) that the modulus of the
spectral densities are bounded above
\begin{equation}
\label{boundab} |{\varrho}_{n}^{(2)}(\varsigma)|<({\theta}_{\rm
max})^{n},
\end{equation}
where ${\theta}_{\rm max}={\cal R}_{\rm max}/b_{1}$ and ${\cal
R}_{\rm max}$ is the maximal value of ${\cal R}$ in the range
$0<\varsigma<\infty$. We find useful to use the ``Maple 7"
\cite{corl} for determining ${\cal R}_{\rm max}$ numerically. In
Table 12, we listed numerical values of $\theta_{\rm max}$ in the
phenomenologically interesting case $n_{f}=\{0-6\}$.
\begin{table}
\caption{The quantity $\theta_{\rm max}$ versus the convergence
radii ${\tilde\rho}^{(k)}$.}
\begin{indented}
\item[]\begin{tabular}{@{}llllllll}\hline $n_{\rm f}$& 0& 1& 2& 3&
4& 5&6
\\ \hline
$\theta_{max}$& 0.237& 0.237& 0.238& 0.240& 0.243& 0.249& 0.259\\
${\tilde\rho}^{(3)}$&0.594&0.601&0.618&0.653&0.731&0.956&1.885\\
${\tilde\rho}^{(4)}$& 0.485 &0.488 &0.497&0.516 &0.554 &0.641 &0.779\\
 \hline
\end{tabular}
\end{indented}
\end{table}
Note that all the power series $\sum_{n=m}^{\infty}{\cal
C}_{m,n}^{(k)}{\theta^{n}}$, $m=1,2\ldots$, have the same radius
of convergence, ${\tilde\rho}^{(k)}$, as the original series
(\ref{ser2}). This follows from the definition
\begin{equation}
\label{serm} \sum_{n=m}^{\infty}{\cal
C}_{m,n}^{(k)}{\theta^{n}}=\left(\sum_{l=1}^{\infty}{\tilde
c}_{l}^{(k)}\theta^{l}\right)^{m}.
\end{equation}
Consider now the set of numerical series of positive terms
\begin{equation}
\label{numser} \sum_{n=m}^{\infty}|{\cal
C}_{m,n}^{(k)}|{\theta_{max}^{n}}\quad m=1,2\ldots,
\end{equation}
looking at the numbers in Table 12, we see that $\theta_{\rm max}$
is inside the convergence disk of the series (\ref{serm}),
$0<\theta_{\rm max}<\tilde\rho^{(k)}$. Hence all the numerical
series (\ref{numser}) are
convergent. Combining this  fact with the bounding conditions
(\ref{boundab}), we find that the series of functions
$\sum_{n=m}^{\infty}|{\cal C}_{m,n}^{(k)}\varrho^{(2)}_{n}
(\varsigma)|$, $m=1,2\ldots$, are uniformly convergent by the
comparison test due to Weierstrass. Then all the series
(\ref{rhoser}) are  uniformly convergent. Hence  by the arguments
given above,   the series of functions (\ref{acalser}) and
(\ref{mathser})  are also uniformly convergent.

The series (\ref{acalser}) and (\ref{mathser}) enable us to
calculate the infrared limits of the APT expansion functions. Thus
we may  reproduce remarkable results of Shirkov and Solovtsov
\cite{ss,ss1} in a mathematically rigorous way \footnote[5]{An
alternative derivation of these results in the context of the
asymptotic solutions to the RG equation was recently given  in
ref.~\cite{AA1}.}. It is seen  from the definition (\ref{spectrl})
that $\lim_{u\rightarrow 0^{+}}{\cal A}^{(k)}_{n}(u)=\lim_{{\bar
s}\rightarrow 0^{+}}{\mathfrak A}^{(k)}_{n}({\bar s})$. Therefore,
we shall consider only the Minkowskian set of functions. In the
2-loop order, one may calculate the infrared limits of the
expansion functions using the explicit formulas obtained in
ref.~\cite{mpr}. The first two  Minkowskian functions are given by
\begin{equation}
\label{mink2} \fl {\mathfrak A}_{1}^{(2)}({\bar s})=1-{\pi}^{-1}
\Im\ln{W_{1}(z_{s})},\quad {\mathfrak A}_{2}^{(2)}({\bar
s})={\pi}^{-1}b_{1}^{-1}{\Im}\ln\{W_{1}(z_{s})/(1+W_{1}(z_{s}))\},
\end{equation}
where $z_{s}=(eb_{1})^{-1}{\bar s}^{-1/b_{1}}\exp(\imath\pi
(1/b_{1}-1))$. The  functions with higher values of index are
determined by the recurrence relation (see ref.~\cite{mpr})
\begin{equation}
\label{rr} {\mathfrak A}_{n+2}^{(2)}({\bar
s})=-b_{1}^{-1}\left({\mathfrak A}_{n+1}^{(2)}({\bar s})+{1\over
n}\frac{\rm d}{{\rm d}\ln{\bar s}}{\mathfrak A}_{n}^{(2)}({\bar
s})\right).
\end{equation}
The asymptotics of $W_{1}(z_{s})$ as $|z_{s}|\rightarrow \infty$
may be determined using Eq.~(\ref{wasy}) with
$L_{1}=\ln{z_{s}}+\imath 2\pi$ \cite{lamb}. Combining
(\ref{mink2}) and (\ref{rr}) and taking the limit $\bar
s\rightarrow 0^{+}$, we find
\begin{equation}
{\mathfrak A}_{n}^{(2)}({\bar s})\approx
{\delta}_{n,1}+(1+b_{1})\ln^{-n}{\bar s}+O(\ln^{-n-1}{\bar
s})\rightarrow \delta_{n,1},
\end{equation}
hence    $\lim_{u\rightarrow 0^{+}}{\cal A}_{n}^{(2)}(u)=
\lim_{{\bar s}\rightarrow 0^{+}}{\mathfrak A}_{n}^{(2)}({\bar
s})={\delta}_{n,1}$. These relations may be extended to higher
orders by means of the expansions (\ref{acalser}) and
(\ref{mathser}). Thus we can write
\begin{equation}
\lim_{u\rightarrow 0^{+}}{\cal A}_{m}^{(k)}(u)=
\sum_{n=m}^{\infty}{\cal C}_{m,n}^{(k)}\lim_{u\rightarrow
0^{+}}{\cal A}_{n}^{(2)}(u)={\cal
C}_{m,1}^{(k)}\equiv\delta_{m,1},
\end{equation}
the calculation of the limit of the sum of the series
term-by-term, as $u\rightarrow 0$, is justified  by the uniform
convergence of the series. It should be stressed that these
results (in particular, finiteness of ${\cal A}_{1}^{(k)}(0)$
originally obtained in \cite{ss}) are direct consequences of the
asymptotic freedom (AF) of the theory. This interesting
relationship has been recently elucidated by the author of
ref.~\cite{AA1} using a different technique. In this connection, we
remark that the recurrence formula (\ref{rr}) follows directly
from the AF, as it was shown in refs.~\cite{mpr,join}. The
universality of ${\cal A}_{1}^{(k)}(0)$ and ${\mathfrak
A}_{1}^{(k)}(0)$ (the scheme independence and invariance with
respect to higher-loop corrections) is evident.

\section{Conclusion}

The main objective of this investigation was to study convergence
properties of the new expansion (\ref{ser}).  In section 2 we have
systematically discussed  the analyticity structure of the modified coupling $a(Q^2/\Lambda^2)$
at 3- and 4-loops in the complex $u=Q^2/\Lambda^2$ plane for all
$n_{\rm f}$ values in the range $0\leq n_{f}\leq 16$.  For higher
values of $n_{f}$, when the $\beta$-function has only real zeros,
we have reproduced the part of results of refs.~\cite{gg,gkarl}
using a different technique. For low values of $n_{f}$, when the
$\beta$-function has complex zeros, we have determined the
analytical continuation of the function $t=T(a)$ choosing the cuts properly
in the complex coupling plane. With this choice, we have found that the running
coupling has a pair of complex conjugate singular points in the
first Riemann sheet of the $Q^2$ plane besides the real
singularity on the positive semi-axis. In many cases, just these
complex singularities determine the radius of convergence of the
series (\ref{ser}) (e.g. $0\leq n_{f}\leq 5$ at 3 loops).

In section 3 we have proved that in the $\overline {{\rm
MS}}$-like  schemes the power series (\ref{ser}) has a finite
radius of convergence to all orders in perturbation theory for all
$n_{f}=0-16$. Therefore, the series inside its circle of
convergence represents the exact solution to the RG equation
(\ref{eff}). In the proof, we have used methods of analytical
theory of differential equations.

In section 4 we have determined the analytical structure of the
modified coupling in higher orders as a function of the 2-loop
order coupling $\theta$ ($\theta=a^{(2)}(Q^2/\Lambda^2)$). We have
considered the 3- and 4-loop cases for $0\leq n_{f}\leq 16$ and
$0\leq n_{f}\leq 7$ respectively. We have found helpful
Eq.~(\ref{wsol1}), the implicit solution for the higher order
coupling determined via the Lambert-W function. By means of this
formula, we have determined the analytical continuation of the
inverse function $\theta=F^{-1}(a)$ in the complex $a$ plane. This
enabled us to find the locations of the singularities of the
coupling $a=F(\theta)$ in the $\theta$-plane (see Tables 6 and 7).
The correspondence between the singularities of the coupling in
the $Q^2$ and $\theta$ planes has been established. Comparing
various singularities of the coupling in the $\theta$-plane, we
have determined the radii of convergence of the series
(\ref{ser2}) $\tilde{\rho}^{(k)}$.
From a practical viewpoint, the radii of
convergence of the original series (\ref{ser}), $\rho^{(k)}=\tilde{\rho}^{(k)}/\beta_{0}$, is found to be
sufficiently large.
For example, ${\rho}^{(3)}=0.965$ and
${\rho}^{(4)}=0.720$ at $n_f=3$. The obtained predictions for the
convergence radius have been examined by the independent numerical
calculation. One further important property of the series is high
convergence rate. In previous papers \cite{join,m4}, we observed
that in the 3- and 4-loop orders partial sums of these series with
the first few terms give very good approximations to the coupling
even in the infrared region. This was confirmed in conventional
perturbation theory as well as in APT.

In section 5 we have determined the convergence region of the
series (\ref{ser2}) in the momentum squared space. For
sufficiently large $n_{f}$ values ($n_{\rm f}\geq 14$ in the
${\overline{\rm MS}}$ scheme), we have found that the series
converges in the whole physical range $0<Q^2<\infty$. For  lower
$n_{\rm f}$ values, we have evaluated the lower boundary of the
convergence region $Q^{2}_{\rm min}$. We have compared this scale
with the estimations of the infrared boundary of QCD, $\mu_{\rm c}$,  obtained
within two different non-perturbative approaches and found that
$Q^{2}_{\rm min}<\mu_{\rm c}^{2}$. This is in agreement with the  possibility
that the series (\ref{ser}) in the ${\overline{\rm MS}}$ scheme
may be used safely in the whole perturbative region $\mu_{\rm
c}^{2}<Q^2<0$.

In section 6 we have studied the convergence properties of the
non-power series constructed from the series (\ref{ser2})
according to the rules  of the QCD Analytic Perturbation Theory of
Shirkov and Solovstov both in  the space- and time-like regions. We
have shown that the Euclidean and Minkowskian variants of these
non-power series are uniformly convergent   over whole  domains of
the corresponding momentum squared variables. A mathematically
rigorous proof of an interesting
result of ref.~\cite{ss}, the finiteness and universality of the analytic
coupling at zero momentum, has been presented.

The series solution (\ref{ser}) may be useful in high-precision
calculations of QCD observables beyond the 2-loop order in
low-momentum regime. It clearly  provides more accurate results
than the standard asymptotic expansion (\ref{asycoup}) for the coupling
 (see ref.~\cite{m4}).  This series  may be used in
different variants of the analytic approach to perturbative QCD
suggested in refs.~\cite{mss, AA, cvetic,bakul}. It may also be
applied in the contexts of the ``contour improved" perturbation
theory of refs.~[26-31] and resummation methods proposed in
refs.~\cite{mx1} and \cite{cvetic1}. Another possible application
of the series is to construct the running coupling with consistent
matching conditions at quark thresholds in MS-like renormalization
schemes
 \cite{shirk2, AA1, mpr,rdr}.

\ack{I would like to thank D. V. Shirkov for helpful advises. The
author have enjoyed discussions on this subject with  A. L.
Kataev, A. A. Khelashvili, A. N. Kvinikhidze, S.~V. Mikhailov and
I. L. Solovtsov. The present work has been partly supported by the
Georgian Research and Development Foundation under grant No,
GEP2-3329-TB-03.}

\appendix
\section*{Appendix}
\setcounter{section}{1}
\setcounter{equation}{0}
The RG equation to the $k$-th order reads
\begin{equation}
\label{eff} {{d}\alpha_{\rm{s}}(Q^2)\over{d}\ln {Q^{2}}}=
{\beta}^{(k)}(\alpha_{\rm{s}}(Q^2))=-\sum_{n=0}^{k-1}
\beta_{n}\{\alpha_{\rm{s}}(Q^2)\}^{n+2},
\end{equation}
the running coupling satisfies the normalization condition
$\alpha_{\rm s}(\mu^{2})=g^2/(4\pi)$, where $\mu$ is the
renormalization point and $g$ is the gauge coupling of QCD. In the
class of schemes where the $\beta$-function is mass independent
$\beta_0$ and $\beta_1$ are universal
\begin{equation}
\label{coef} \beta_{0}=(4\pi)^{-1}(11-2 n_{\rm f}/3),\qquad
\beta_{1}=(4\pi)^{-2}(102-38 n_{\rm f}/3).
\end{equation}
The results for the coefficients $\beta_{2}$ and $\beta_{3}$ in
the $\overline{MS}$ scheme can be found in refs. \cite{tvz} and
\cite{rvl}
\begin{equation}
\beta_{2}=(4\pi)^{-3}(2857/2-5033n_{\rm f}/18+325n_{\rm f}^2/54)
\end{equation}
\begin{eqnarray}
\beta_{3}=(4\pi)^{-4}\left(\frac{149753}{6}+3564\zeta_{3}-\left(\frac{1078361}{162}
+\frac{6508}{27}\zeta_{3}\right)n_{\rm f}\right.\nonumber\\
+\left.\left(\frac{50065}{162}+\frac{6472}{81}\zeta_{3}\right)n_{\rm
f}^{2}+\frac{1093}{729}n_{\rm f}^{3}\right),
\end{eqnarray}
here $\zeta$ is the Riemann zeta-function
($\zeta_{3}=1.202056903...$). The values of the first three
coefficients $b_{1,2,3}$ ($b_{n}=\beta_{n}/\beta_{0}^{n+1}$) in
the ${\overline{{\rm MS}}}$ scheme are tabulated in Table A.1.
\begin{table}
\caption{The ${\overline{{\rm MS}}}$-scheme ${\bar\beta}$ function
coefficients $b_{1, 2, 3}$ for $n_{f}=0-16$.}
\begin{tabular}{@{}llllcllll}
\hline $ n_{f}$ & $b_{1}$ &
$b_{2}$& $b_{3}$& &$n_{f}$&$b_{1}$ & $b_{2}$ &$b_{3}$\\
\cline{1-4}\cline{6-9}

0&102/121 &2857/2662 &1.9973 & &9 &-12/25 &-1201/250&1.0105 \\
1& 804/961& 62365/59582& 1.9913& & 10 &-222/169& -41351/4394& 5.0716\\

2&690/841 &48241/48778 &1.9449 & &11 &-336/121 &-49625/2662 &21.273\\

3&64/81& 3863/4374 &1.8428 && 12 &-50/9 &-6361/162 &84.088 \\

4& 462/625& 21943/31250 & 1.6662& &13 &-564/49 &-64223/686 & 360.81\\

5&348/529 &9769/24334 &1.3969 & &14 &-678/25 &-70547/250 &2009.6 \\

6& 26/49 &-65/686 &1.0297 & &15 &-88 &-2823/2 &21254\\

7& 120/361 &-12629/13718 & 0.6107& &16 &-906 &-81245/2 &2263651\\

8& 6/289 &-22853/9826 &0.3549 & & & & &\\

\hline
\end{tabular}
\end{table}

\end{document}